\documentclass[preprint2]{aastex631}

\shorttitle{PHIBSS Outflows}
\shortauthors{Barfety et al.}
\graphicspath{{./}{figures/}}

\begin{document}

\title{PHIBSS: Searching for Molecular Gas Outflows in Star-Forming Galaxies at $z = 0.5 - 2.6$}

\author[0000-0002-1952-3966]{Capucine Barfety}
\affiliation{Max-Planck-Institut für extraterrestrische Physik, Giessenbachstrasse, D-85748 Garching, Germany}

\author[0000-0002-3405-5646]{Jean-Baptiste Jolly}
\affiliation{Max-Planck-Institut für extraterrestrische Physik, Giessenbachstrasse, D-85748 Garching, Germany}

\author[0000-0003-4264-3381]{Natascha M. F\"orster Schreiber}
\affiliation{Max-Planck-Institut für extraterrestrische Physik, Giessenbachstrasse, D-85748 Garching, Germany}

\author[0000-0002-1485-9401]{Linda J. Tacconi}
\affiliation{Max-Planck-Institut für extraterrestrische Physik, Giessenbachstrasse, D-85748 Garching, Germany}

\author[0000-0002-2767-9653]{Reinhard Genzel}
\affiliation{Max-Planck-Institut für extraterrestrische Physik, Giessenbachstrasse, D-85748 Garching, Germany}
\affiliation{Departments of Physics and Astronomy, University of California, Berkeley, CA 94720, USA}

\author[0000-0003-4226-7777]{Giulia Tozzi}
\affiliation{Max-Planck-Institut für extraterrestrische Physik, Giessenbachstrasse, D-85748 Garching, Germany}

\author[0000-0001-6879-9822]{Andreas Burkert}
\affiliation{Max-Planck-Institut für extraterrestrische Physik, Giessenbachstrasse, D-85748 Garching, Germany}
\affiliation{University Observatory Munich (USM) Scheinerstrasse 1, D-81679 Munich, Germany}

\author[0000-0003-3921-3313]{Jianhang Chen}
\affiliation{Max-Planck-Institut für extraterrestrische Physik, Giessenbachstrasse, D-85748 Garching, Germany}

\author[0000-0003-2658-7893]{Françoise Combes}
\affiliation{Observatoire de Paris, LUX, CNRS, PSL Univ., Sorbonne University, Paris, France}
\affiliation{College de France, 11 Pl. Marcelin Berthelot, 75231 Paris, France}

\author[0000-0003-4949-7217]{Ric Davies}
\affiliation{Max-Planck-Institut für extraterrestrische Physik, Giessenbachstrasse, D-85748 Garching, Germany}

\author{Frank Eisenhauer}
\affiliation{Max-Planck-Institut für extraterrestrische Physik, Giessenbachstrasse, D-85748 Garching, Germany}
\affiliation{Department of Physics, Technical University of Munich, 85748 Garching, Germany}

\author[0000-0001-6703-4676]{Juan M. Espejo Salcedo}
\affiliation{Max-Planck-Institut für extraterrestrische Physik, Giessenbachstrasse, D-85748 Garching, Germany}

\author[0000-0002-2775-0595]{Rodrigo Herrera-Camus}
\affiliation{Departamento de Astronomía, Universidad de Concepción, Barrio Universitario, Concepción, Chile}

\author[0000-0001-7457-4371]{Lilian L. Lee}
\affiliation{Max-Planck-Institut für extraterrestrische Physik, Giessenbachstrasse, D-85748 Garching, Germany}

\author[0000-0002-2419-3068]{Minju M. Lee}
\affiliation{Cosmic Dawn Center (DAWN), Denmark}
\affiliation{DTU-Space, Technical University of Denmark, Elektrovej 327, DK2800 Kgs. Lyngby, Denmark}

\author[0000-0001-9773-7479]{Daizhong Liu}
\affiliation{Max-Planck-Institut für extraterrestrische Physik, Giessenbachstrasse, D-85748 Garching, Germany}
\affiliation{Purple Mountain Observatory, Chinese Academy of Sciences, 10 Yuanhua Road, Nanjing 210023, People’s Republic of China}

\author[0000-0003-0291-9582]{Dieter Lutz}
\affiliation{Max-Planck-Institut für extraterrestrische Physik, Giessenbachstrasse, D-85748 Garching, Germany}

\author[0000-0002-7314-2558]{Thorsten Naab}
\affiliation{Max-Planck Institute for Astrophysics, Karl Schwarzschildstrasse 1, 85748, Garching, Germany}

\author[0000-0002-7176-4046]{Roberto Neri}
\affiliation{Institut de Radioastronomie Millim\'etrique (IRAM), 300 Rue de la Piscine, 38400 Saint-Martin-d’H\`eres, France}

\author[0000-0003-1785-1357]{Amit Nestor Shachar}
\affiliation{School of Physics and Astronomy, Tel Aviv University, Tel Aviv 69978, Israel}

\author[0009-0009-0472-6080]{Stavros Pastras}
\affiliation{Max-Planck-Institut für extraterrestrische Physik, Giessenbachstrasse, D-85748 Garching, Germany}
\affiliation{Max-Planck Institute for Astrophysics, Karl Schwarzschildstrasse 1, 85748, Garching, Germany}

\author[0000-0002-1428-1558]{Claudia Pulsoni}
\affiliation{Max-Planck-Institut für extraterrestrische Physik, Giessenbachstrasse, D-85748 Garching, Germany}

\author[0000-0002-0108-4176]{Sedona H. Price}
\affiliation{Department of Physics and Astronomy and PITT PACC, University of Pittsburgh, Pittsburgh, PA 15260, USA}

\author[0000-0002-7093-7355]{Alvio Renzini}
\affiliation{Osservatorio Astronomico di Padova, Vicolo dell’Osservatorio 5, Padova, I-35122, Italy}

\author{Karl Schuster}
\affiliation{Institut de Radioastronomie Millim\'etrique (IRAM), 300 Rue de la Piscine, 38400 Saint-Martin-d’H\`eres, France}

\author[0000-0002-2125-4670]{Taro T. Shimizu}
\affiliation{Max-Planck-Institut für extraterrestrische Physik, Giessenbachstrasse, D-85748 Garching, Germany}

\author[0000-0001-5065-9530]{Amiel Sternberg}
\affiliation{School of Physics and Astronomy, Tel Aviv University, Tel Aviv 69978, Israel}

\author[0000-0002-0018-3666]{Eckhard Sturm}
\affiliation{Max-Planck-Institut für extraterrestrische Physik, Giessenbachstrasse, D-85748 Garching, Germany}

\author[0000-0003-4891-0794]{Hannah \"Ubler}
\affiliation{Max-Planck-Institut für extraterrestrische Physik, Giessenbachstrasse, D-85748 Garching, Germany}

\author[0000-0003-3735-1931]{Stijn Wuyts}
\affiliation{ Department of Physics, University of Bath, Claverton Down, Bath, BA2 7AY, UK}



\begin{abstract}

We present an analysis of millimeter CO observations to search and quantify signatures of molecular gas outflows. We exploit the large sample of $0.5<z<2.6$ galaxies observed as part of the PHIBSS1/2 surveys with the IRAM Plateau de Bure interferometer, focusing on the 154 typical massive star-forming galaxies with CO detections (mainly CO(3-2), but including also CO(2-1) and CO(6-5)) at signal-to-noise (SNR) $>$ 1.5 and available properties (stellar mass, star formation rate, size) from ancillary data. None of the individual spectra exhibit a compelling signature of CO outflow emission even at high SNR $>$ 7. To search for fainter outflow signatures, we carry out an analysis of stacked spectra, including the full sample, as well as subsets, split in terms of stellar mass, redshift, inclination, offset in star formation rate (SFR) from the main sequence, and AGN activity. None of the physically motivated subsamples show any outflow signature. We report a tentative detection in a subset statistically designed to maximize outflow signatures. We derive upper limits on molecular gas outflow rate and mass loading factors $\eta$ based on our results and find $\eta \leq 2.2 - 35.4$, depending on the subsample. Much deeper CO data and observations of alternative tracers are needed to decisively constrain the importance of cold molecular gas component of outflows relative to other gas phases.


\end{abstract}



\section{Introduction} \label{sec:intro}

Feedback in the form of outflows has long been invoked to explain observed galaxy scaling relations and stages of galaxy evolution. They are believed to be a key part of the baryon cycle, mixing and redistributing gas within and around galaxies \citep{tumlinson17, peroux2020}. As such, they are important mechanisms to shape the mass-metallicity relation, set metallicity gradients within galaxies \citep{dave11, sanders18}, and explain the large reservoirs of baryons and metals in the intergalactic medium (IGM) \citep{peeples14,tumlinson17}. They are also invoked to reconcile theoretical and numerical predictions with observations, such as the galaxy mass function (e.g., \citealt{schaye15}), the color bi-modality (blue, young star-forming galaxies (SFGs) vs. red passive galaxies), or even scaling relations between black hole mass and host galaxy bulge properties \citep{dekel1986,mutch2013}.

Outflows originate from two processes: stellar winds and supernovae in star-forming regions \citep{murray2005, hopkins14} and active galactic nuclei (AGN) accretion \citep{fabian2012, king2015}. Depending on their driving mechanism, the outflow properties can vary strongly. Star formation (SF)-driven outflows have velocities $\sim 10^2$ km/s, unlikely to escape the potential well of the galaxy (e.g., \citealt{shapley2003, newman2012b, leroy2015, fs19, rebecca2019, avery2021}), regulating SF and galaxy growth over the span of galaxies' life on and above the main sequence (MS; \citealt{rc2025}). On the other hand, AGN-driven outflows happen over shorter timescales and are typically more powerful, driving faster winds ($\gtrsim 10^3$ km/s) reaching further into the circumgalactic medium (\citealt{rebecca2020, hc19}; see \citealt{harrison2024} for a recent review). Defining the mass loading factor $\eta = \frac{\dot{M}_{\mathrm{out}}}{\mathrm{SFR}}$ as a measure of the dominant gas depletion mechanism, where $\eta > 1$ is interpreted as outflows carrying enough mass to deplete the gas faster than SF can, AGN-driven outflows typically have higher mass loading factors than SF-driven outflows, and thus affect SF more efficiently \citep{fiore2017}.

At $z = 1 - 3$, the cosmic star formation rate (SFR) and AGN density reach their peak, and so does the cold molecular gas fraction in galaxies \citep{madau2014, tacconi2020}. During this ``cosmic noon'' epoch, roughly half of the present-day stellar mass is formed, most of which ($\sim 90$\%) is taking place in SFGs on/near the MS \citep{rodighiero2011, speagle2014, whitaker2014}. Studying outflows in MS SFGs at $z \sim 1 - 3$ is important to quantify the impact of SF and AGN feedback on the growth and evolution of the galaxy population as a whole. Observational studies have reported that galactic scale outflows are ubiquitous at this epoch (see \citealt{fs2020} and \citealt{veilleux2020} for recent reviews). At these redshifts, most outflow studies are in the ionized gas phase, detected both through rest-UV interstellar features (e.g., \citealt{shapley2003, talia2012, talia2017, weldon2022, calabro2022, kehoe2024}) and through optical emission lines such as H$\alpha$, H$\beta$, [OIII] and [NII] (e.g., \citealt{genzel2011, genzel2014, harrison2016, leung2017, leung2019, swinbank2019, fs19, rebecca2020, concas2022, weldon2024, kehoe2024}). This body of work highlighted correlations of outflow incidence with stellar mass, star formation rate (SFR), distance from the MS ($\Delta$MS), and redshift. However, looking into the mass loading factor $\eta$, studies find typical values of $\eta < 1$ for both AGN- and SF-driven ionized gas outflows (e.g., \citealt{freeman19, fs19, rebecca2020}). These low values of $\eta$ for ionized gas outflows are in contradiction with the scenario where outflows are believed to regulate SF on galactic scales.


These results at cosmic noon are unsurprising as the ionized gas phase may not carry the bulk of the mass (although they carry substantial energy and momentum; e.g., \citealt{fs19}). At all redshifts, observations show that outflows are complex and multi-phase, with observations across the whole wavelength range spanning wide physical scales \citep{genzel2011, genzel2014, bolatto2013, fluetsch19, hc2020, levy2021, rc2022, tozzi2021, kehoe2024, rebecca24, parlanti2024}, a result corroborated by simulations \citep{cooper2008, costa2015, schneider2018, ward2024}. Observations of the cold molecular phase of outflows in local galaxies, through CO transition lines or P-Cygni profiles in far-infrared (IR) OH lines, find that they are much more efficient drivers of material than their ionized gas counterparts \citep{sturm2011, veilleux2013, bolatto2013b, contursi2013, ric2014, ga2017, krieger2019, schneider2020, vijayan2024a}. Except for one AGN-hosting MS galaxy at $z > 2$ \citep{hc19}, detections of molecular gas outflows at cosmic noon have been exclusively in luminous AGNs and quasars (e.g., \citealt{brusa2018, chartas2020, rebecca2020}), which may not be representative of the general population of galaxies. Another stacking analysis of CO observations of cosmic noon MS and starbursting galaxies find no statistically significant outflow detection (Langan et al. in prep). This is partly due to the challenging nature of CO observations at higher redshift; however, in this framework, their incidence and properties among cosmic noon galaxies and their impact on galaxy evolution have yet to be established.

This paper aims to address this outstanding issue by exploiting the PHIBSS 1/2 datasets of mostly main-sequence $z = 0.5 - 2.6$ SFGs to search for molecular gas outflow signatures in CO mid-\textit{J} transitions. The sample is unbiased towards AGN activity and offers a population-averaged view of typical SFGs at cosmic noon, and is the largest, most complete sample available to conduct this analysis at cosmic noon. In Section \ref{sec:obs}, we introduce the survey and galaxy sample used in the analysis, as well as the details of the observations. In Sections \ref{sec:methods} and \ref{sec:res}, we present the details of the analysis and results of individual and stacked galaxy spectra. We present the tests of the method in Section \ref{sec:test}. We discuss the implications of our results in Section \ref{sec:disc}, and summarize our conclusions in Section \ref{sec:conc}. Throughout the paper we use a standard $\Lambda$CDM cosmology with H$_0$ = 70 km s$^{-1}$ Mpc$^{-1}$, $\Omega_m = 0.3$, and $\Omega_\Lambda = 0.7$.

\section{Observations}\label{sec:obs}
\subsection{The PHIBSS sample}\label{sec:phibss}

The Plateau de Bure High-\textit{z} Blue Sequence Survey (PHIBSS) is a molecular gas survey of 174 typical MS galaxies spanning $0.5 \le z \le 2.6$. Observations were carried out in two large programs, PHIBSS1 (2009-2011; PIs: L. Tacconi \& F. Combes) and PHIBSS2 (2013-2017; PIs: F. Combes, S. García-Burillo, R. Neri \& L. Tacconi), using the Plateau de Bure Interferometer (PdBI)/NOEMA. Observations cover a range of beam sizes from $0.65 - 9.4" \times 0.65 - 5.25"$, with spectral resolutions from 7 km/s to 88 km/s. For all sources, the galaxy properties were derived using ancillary data, mainly \textit{HST}/WFC3 and ACS, using SED modeling to get the stellar masses, and Sersic profile fitting to estimate the effective radii. The SFRs were estimated using UV and IR luminosities, or extinction-corrected H$\alpha$ luminosities for the highest redshift sources. In all cases, the photometry comes from apertures encompassing the flux from the entirety of the galaxy. Similarly, we extract the CO flux from apertures designed to cover the whole galaxy for the majority of sources. Thus, all our measurements should reflect the global galaxy properties. Detailed descriptions of the observations, sample, and galaxy properties' estimations can be found in \cite{tacconi2013, tacconi2018} and \cite{freundlich2019}.

\begin{figure}
    \centering
    \includegraphics[width=\linewidth]{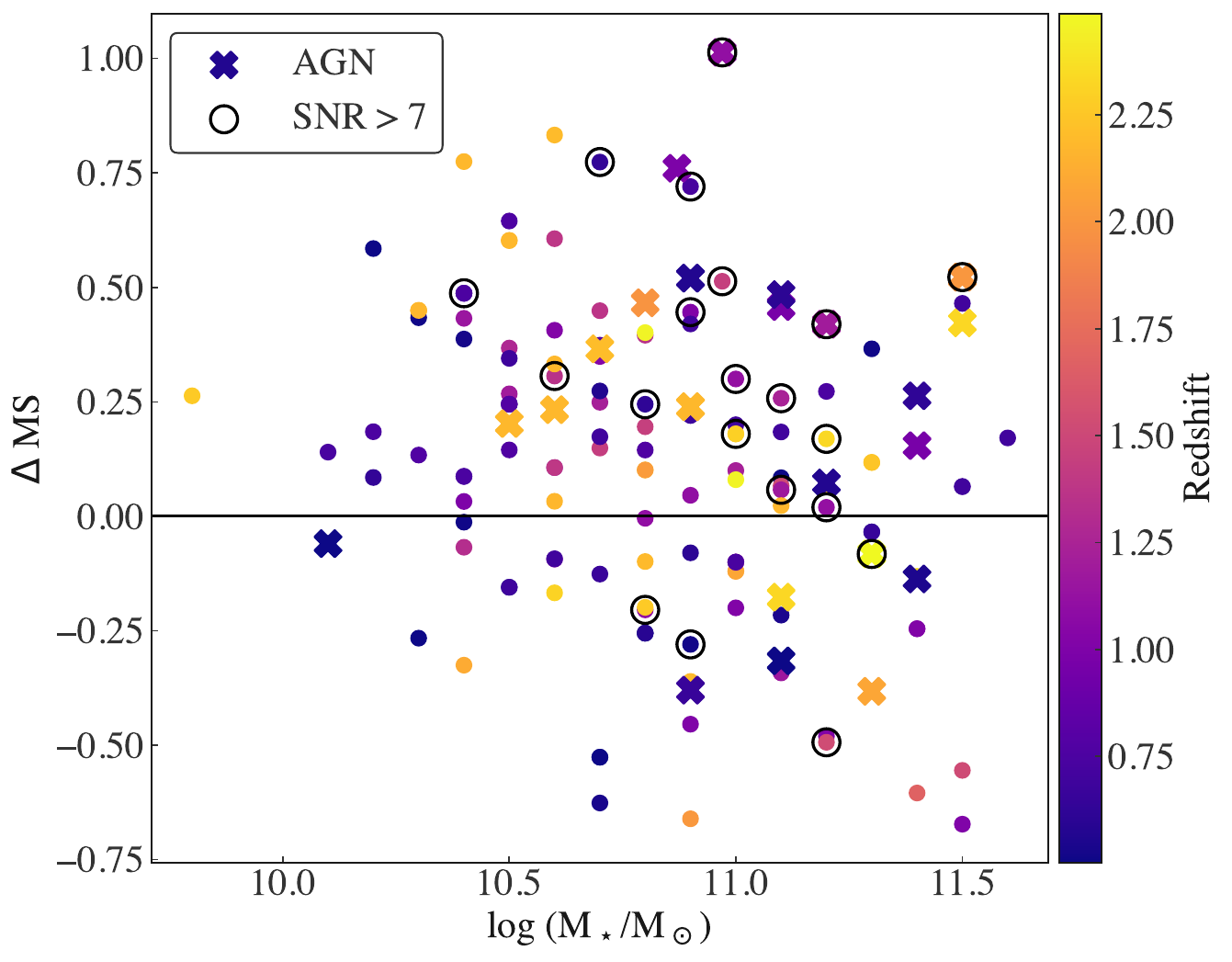}
    \caption{Distribution of the PHIBSS sample used in this paper in distance from the Main Sequence (MS) following the parametrization from \cite{whitaker2014}, as a function of stellar mass. Circled in solid black are the 20 galaxies detected with SNR $> 7$ (the full distribution of SNRs is shown in Appendix \ref{ap:sample}, Fig. \ref{fig:sample_prop}), and galaxies marked with a cross are flagged as AGN-hosting.}
    \label{fig:sample}
\end{figure}

The observations cover the CO(3-2) line for 99 galaxies at redshift $1.00 \leq z \leq 2.55$, the CO(2-1) line for 70 lower redshift ($0.50 \leq z \leq 1.53$) galaxies, and the CO(6-5) line for the remaining 5 galaxies ($2.01 \leq z \leq 2.33$). For each galaxy, we compute the integrated SNR of the narrow emission line and discard detections below SNR = 1.5. Although this threshold is lower than what is typically considered significant, the stacking process aims to uncover the signal hidden within the noise, so we retain these detections in the stack. This process brings the sample down to 154 galaxies, with median $\langle z \rangle = 1.04$, $\langle \mathrm{log(M_\star/M_\odot)} \rangle = 10.8$, $\langle \mathrm{R_e} \rangle = 4.4$ kpc and $\langle \Delta\mathrm{MS} \rangle = 0.14$ dex (see Table \ref{tab:avg_prop}, and the corresponding property distributions are plotted in Appendix \ref{ap:sample}, Fig. \ref{fig:sample_prop}). The individual galaxy properties are reported in Table \ref{tab:sample}, in Appendix \ref{ap:sample}. The distribution is plotted in Fig. \ref{fig:sample} as a function of their distance from the MS (computed following the parametrization from \citealt{whitaker2014}), stellar mass, and redshift. In addition, we verified that the SNR cut does not change our results or improve the SNR of the stacked spectra by repeating the cut with higher threshold values (SNR $> 3$ and $> 5$) and performing the same analysis as described in Sec. \ref{sec:methods}, finding that using higher SNR thresholds yields the same results.

\subsection{AGN Identification}\label{sec:agn}
Since AGNs can be the main driver of strong outflows in the stellar mass range covered by PHIBSS (e.g., \citealt{fs19, rebecca24}), we identify the AGNs among the sample using publicly available multi-wavelength catalogs. We classify the galaxies as AGN based on X-ray, mid-IR, optical, and radio flux diagnostics. 

First, we cross-correlate the PHIBSS catalogs with the \textit{Chandra} X-ray source catalog \footnote{https://cxc.harvard.edu/csc/index.html}. We then apply a luminosity cut based on the galaxies' expected $2 - 10$ keV X-ray luminosity from their SFR \citep{symeonidis2014}, such that galaxies with an associated X-ray detection 1 dex higher than the expected luminosity purely from star formation are flagged as AGNs. From the full sample, 136 galaxies are in \textit{Chandra} coverage, 24 of which have an associated X-ray detection. Of these 24, 17 fulfill the aforementioned criteria. Next, we cross-match the PHIBSS survey with the MOSDEF public line-emission catalogs \citep{kriek2015, reddy2015, shapley2015}, finding 15 galaxies in both samples. We investigate the [OIII]$\lambda$5007/H$\beta$ vs [NII]$\lambda$6584/H$\alpha$ BPT diagram \citep{baldwin1981}, using the upper limit star-forming abundance sequence reported by \cite{kewley2013} at the average redshift $\overline{z} = 2.15$ of the 15 galaxies. We find 6 galaxies lying above the limit, which are thus classified as AGN. Following this, we use the \textit{Spitzer}/IRAC photometry available for 116 of the galaxies to identify AGN based on the color criteria presented by \cite{donley2012}. Eight of the galaxies are classified as AGN based on their mid-IR colors. Lastly, we cross-match the sample with public source catalogs from JVLA radio surveys. Following \cite{delvecchio2017}, we identify as AGNs sources with L$_{1.4\mathrm{GHz}}$ exceeding the value expected from the far-IR-radio scaling relation for star-forming galaxies. Seven of the PHIBSS galaxies satisfy this criterion. In total, 24 galaxies fulfill one or more of the above criteria. The galaxies are reported as AGN in Table \ref{tab:sample}. We note that due to the sample selection aiming for typical star-forming galaxies, none of the sources in our sample are luminous AGNs or have emission dominated by AGN contribution.

\section{Methods}\label{sec:methods}
Our analysis ultimately uses spectral stacking to uncover outflow signatures in the line profile. The stacking process applied in this analysis consists of averaging spectra with potential underlying outflow emission, with various widths, amplitude, and central velocities. In these conditions, we expect outflow signatures in the stacked spectra to appear as a secondary faint broad component centered close to a brighter, narrower emission line originating from gravitationally bound material in the galaxies. This section describes the procedure we apply to retrieve such outflow signatures adequately.

\subsection{Spectrum Extraction}\label{sec:spec}
We use the following method to extract the spectra of individual galaxies and maximize the SNR. First, we define a beam-sized aperture at the center of the data cube, sum the spectra of each pixel within the aperture, and perform a Gaussian fit to the line profile. Using the fit results, we collapse the initial data cubes along the velocity channels, summing all the velocity bins within the FWHM of the line and centering on the best-fit velocity. We then run {\scshape SExtractor} \citep{1996A&AS..117..393B} on the collapsed image to determine the area of detection of the galaxy and create a spatial mask of this detection. Since, in most cases, the detected area by {\scshape SExtractor} is smaller than the beam size, we use the beam size as the aperture for the mask, which we center on the {\scshape SExtractor} coordinates. When the detected area is larger than the beam, we use this as the aperture for the mask. In most cases, the source lies at the center of the field of view, and is not resolved above the beamsize. However, for some sources, the iterative process helps improve the detection substantially, by either identifying the correct source position, or assessing the detection area properly. An example of such case is shown Appendix \ref{ap:spec_extr}, where the source is slightly offset from the center of the frame, and Fig. \ref{fig:l14eg002} shows the initial extracted spectrum compared to the spectrum extracted once correcting for the source position, clearly showcasing the difference in detection.

Finally, we sum all non-masked pixels in the 3D cube and fit the resulting spectrum with a single Gaussian. We iterate over this process three times to obtain the best possible mask (i.e.: the mask for which the integrated spectrum returns the line with the highest SNR). Iterating over this process optimizes the detection area and the estimates of the central velocity and FWHM, which are both crucial to our stacking analysis for precise alignment and to correctly re-bin the spectra (see Sec. \ref{sec:re-binning}). The final apertures have median major axis $\langle a \rangle = 25^{+14}_{-10}$ kpc and minor axis $\langle b \rangle = 19^{+11}_{-6}$ kpc. The final FWHMs are reported in Table \ref{tab:sample}, and their distribution is shown in Appendix \ref{ap:sample}, Fig. \ref{fig:sample_prop}.

After the final iteration, we extract the spectra from the masked cube. We also scale each spectrum's flux by normalizing it to the best-fit amplitude of the Gaussian fit to the emission line, such that the brightest galaxies do not dominate the stacked spectrum, as we are interested in the population average.

    

\subsection{Spectral Re-binning}\label{sec:re-binning}
The spectral stacking of emission lines with different widths, due to the different ranges of projected velocities for different galaxies, can lead to the creation of a broad-like component in the stacked line profile, which can be misinterpreted as an outflow signature in the stacked profile (see, e.g. Sec. \ref{sec:nore-bin}; \citealt{stanley2019} and \citealt{concas2022} for detailed discussions). 

To mitigate this shortcoming of the stacking method, we re-bin the spectral channels of each galaxy spectrum according to their estimated FWHM such that each line would have a FWHM covering the same number of spectral channels. Upon inspection of the distribution of FWHMs in our sample (see Appendix \ref{ap:sample}, Fig. \ref{fig:sample_prop}), we choose a width of 7 channels as the common line width for re-binning, to favor down-sampling but to limit the bin size change to a moderate amount (with a maximum factor of 6)\footnote{We have also repeated the measurements using 9, the median of the FWHM distribution, as the reference value, and observed that it does not affect the results}. After re-binning, the average channel width is 37 $\pm$ 25 km/s. This method impacts the accuracy of the velocity information one can retrieve from the spectra but preserves the line shapes such that any underlying broad component revealed by the stacking could not be attributed to numerical effects. We investigate in Sec. \ref{sec:test} whether re-binning impacts our ability to retrieve real underlying outflows.

\subsection{Spectral Stacking}\label{sec:stack}
After re-binning, we stack the normalized spectra of the individual galaxies using the {\scshape LineStacker} package presented in \citealt{jolly2020}. Given a central velocity for each spectrum (here we use the best-fit velocity retrieved in the spectrum extraction process; see Sec. \ref{sec:spec}), the code aligns the spectra and computes the mean of the flux in each spectral channel, where each spectrum is given the same weight in the mean. We stack both the full sample and the following sub-samples (median properties are listed in Table \ref{tab:avg_prop}): above and below $\mathrm{log(M_\star / M_{\odot})} = 10.7$, above and below $\Delta$MS $= 0.2$, and above and below $z = 1.7$. These choices are motivated by the trends in ionized gas outflow incidence of \citealt{fs19}, corresponding to the stellar mass ($\Delta$MS) above which AGN (SF)-driven outflows become more frequent. We also stack a subsample of galaxies with specific SFR $\mathrm{sSFR} > 0.1 \: \mathrm{Gyr}^{-1}$, using the value of local sSFR for which an increase in outflows is observed \citep{rc2025}. Additionally, we separately stack the sub-sample of 24 AGN-hosting galaxies selected in Sec. \ref{sec:agn}. Finally, since outflow detection is also dependent on galaxy orientation, we create and stack subsamples divided by inclination for the 117 galaxies for which we have the inclination: below 30\textdegree, between 30\textdegree and 60\textdegree, and above 60\textdegree. The stacking results for the full sample, the AGN sample, and the sub-samples are presented and discussed in Sec. \ref{sec:fit}. 

To estimate the noise level in the stacked spectra, we extract a ``blank" spectrum from an empty sky region using the same aperture as used to extract the corresponding galaxy spectrum. After stacking them following the same procedure described above, we take the resulting standard deviation as the noise estimate for each stack. Using this method instead of the noise directly in the spectra allows us to ensure we do not include any ``hidden" signal in the noise estimation.

Finally, we repeat the stacking procedure on the same sample and sub-samples but weigh each spectrum by its noise, such that noisier spectra weigh less in the stack. There are no differences in the results from both stacking prescriptions, and for the rest of the analysis, we use the results from the non-weighted stack.



\section{Results}\label{sec:res}

\subsection{Inspection of Individual Spectra}\label{sec:ind_spec}

As a first step of the analysis, we examine the individual spectra with integrated SNR $> 7$ to search for the presence of outflow signatures in these deeper data sets, as this SNR value indicates robust emission line detection. Out of the full sample of 154 galaxies, 20 have emission lines with SNR above this value (SNR $=7.3 - 20.8$, black circles in Fig. \ref{fig:sample}), four of which are identified as AGNs (all galaxies have coverage by either \textit{Chandra}, MOSDEF, \textit{Spitzer} or VLA, see Sec. \ref{sec:agn}). This subsample of high SNR spectra has median $\langle \Delta\mathrm{MS} \rangle = 0.28$, $\langle z \rangle = 1.16$ and $\langle \mathrm{log(M_\star / M_{\odot})} \rangle = 10.99$ (see Table \ref{tab:sample} and Fig \ref{fig:sample}), thus preferentially probing higher stellar masses ($85\%$ have $\mathrm{log(M_\star / M_{\odot})} > 10.7$), and slightly higher z and $\Delta$MS.

Three of these galaxies exhibit notable asymmetry in their CO line profile. For two of them, the line profile is explained by the fact that the galaxy is well detected over more than one beam and that the mask applied in Sec. \ref{sec:spec} does not cover the full galaxy. This can happen during the spectrum extraction process either if the fitting fails to encompass the full line, or if the aperture misses some flux. In such cases, when creating a collapsed cube to estimate the spatial extent of the detection, we do not sum all of the channels of the line, which results in a truncated detection area, and thus the final extracted spectra will also be missing part of the flux. The standardized iterative process described in Sec. \ref{sec:spec} is designed to mitigate this effect as best as possible, but failed in those two cases of galaxies detected well above the beam size. In this case, the line from the initial mask has an asymmetric shape and becomes the double-peaked profile expected from the galactic rotation after adjusting the mask (see Appendix \ref{ap:ind_spec}, Fig. \ref{fig:5123}). 

For the remaining spectrum, EGS13003805, a z = 1.23 galaxy associated with an AGN, it is difficult to ascertain whether the spectrum shows signs of molecular gas outflows, as the line profile displays signs of asymmetry similar to what is described above (see Appendix \ref{ap:ind_spec}, Fig. \ref{fig:3805}). The issue does not appear to come from the masking. Available \textit{HST}/ACS I- and V-band imaging reveals EGS13003805 as a spiral galaxy \citep{tacconi2013}. In addition, the galaxy is detected over an area larger than the beam size. With this information, and comparing the ALMA detection size and the \textit{HST} size (which are comparable), we favor the interpretation that the double peak in the integrated spectrum is due to galaxy rotation, rather than being an outflow signature, although a more in-depth analysis of the kinematics is required to fully rule out the outflow scenario.


Thus, we observe that prominent outflow signatures are not prevalent at high SNR, even for a subsample with galaxy properties for which we would expect higher outflow incidences and strength (high mass, high redshift, high SFR, AGN hosting). We also investigate later the presence of extended emission beyond the initial masking process (see Sec. \ref{sec:big_ap}). Moreover, we do not expect the masking issue presented above to affect our results more broadly as, in most cases, the detected area is smaller than the beam size, so the majority of our masks encompass more than the detection. 

To fully exploit the PHIBSS data set, we now use the stacked spectra to search for potential outflow signatures that cannot be identified on an individual galaxy basis.


\subsection{Outflow Retrieval from Stacks}\label{sec:fit}
\begin{figure}
    \centering
    \includegraphics[width=\linewidth]{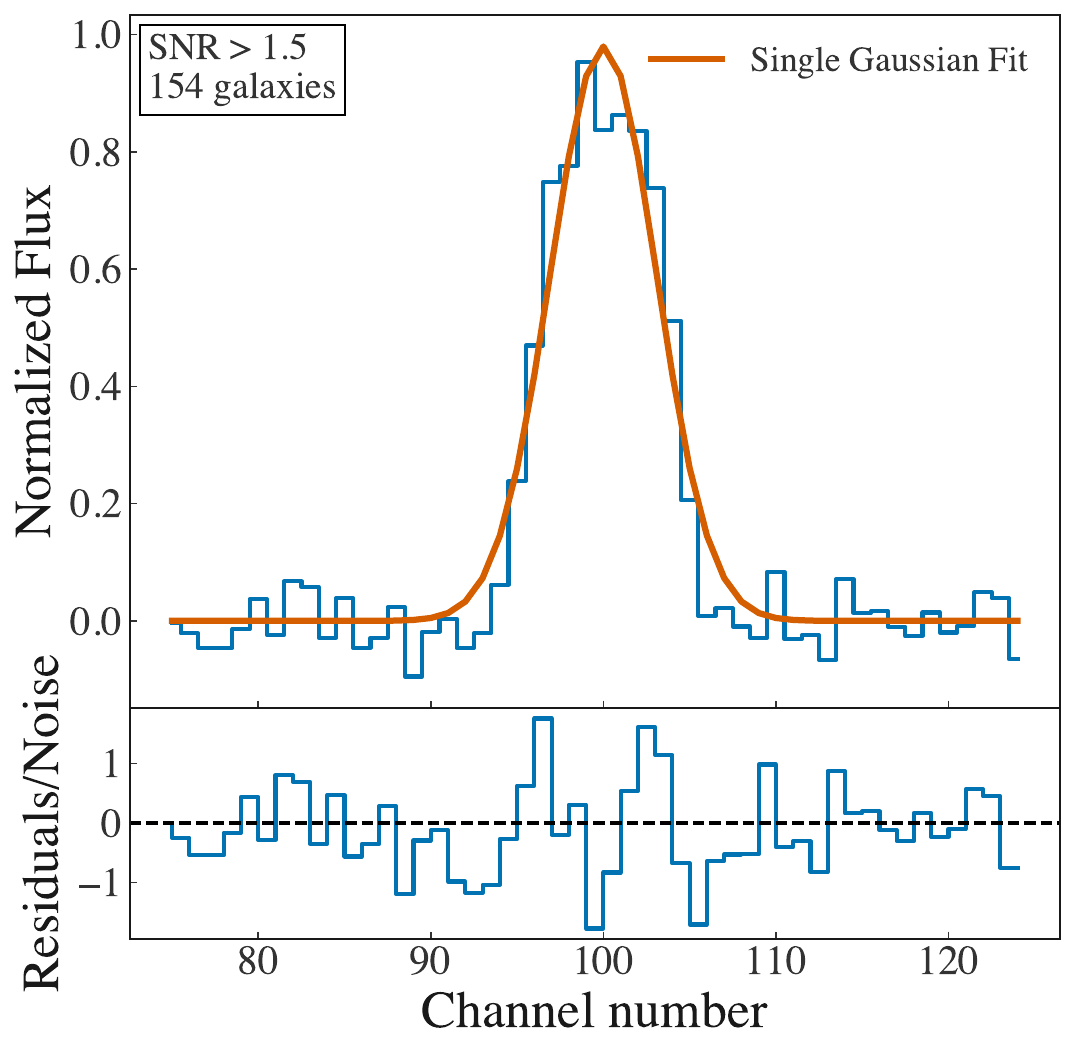}
    \caption{\textit{Upper Panel}: Stacked spectrum (light blue curve) of the full sample of 154 galaxies. Overlaid on the spectrum is the best-fit single Gaussian profile (solid orange). Fitting shows that no significant outflow component is detected in the stacked spectrum. {\textit{Bottom Panel}:} Normalized fit residuals ($\mathrm{\frac{data-model}{noise}}$) from the single Gaussian fitting.}
    \label{fig:full_stack}
\end{figure}

\begin{figure*}
    \centering
    \includegraphics[width=\linewidth]{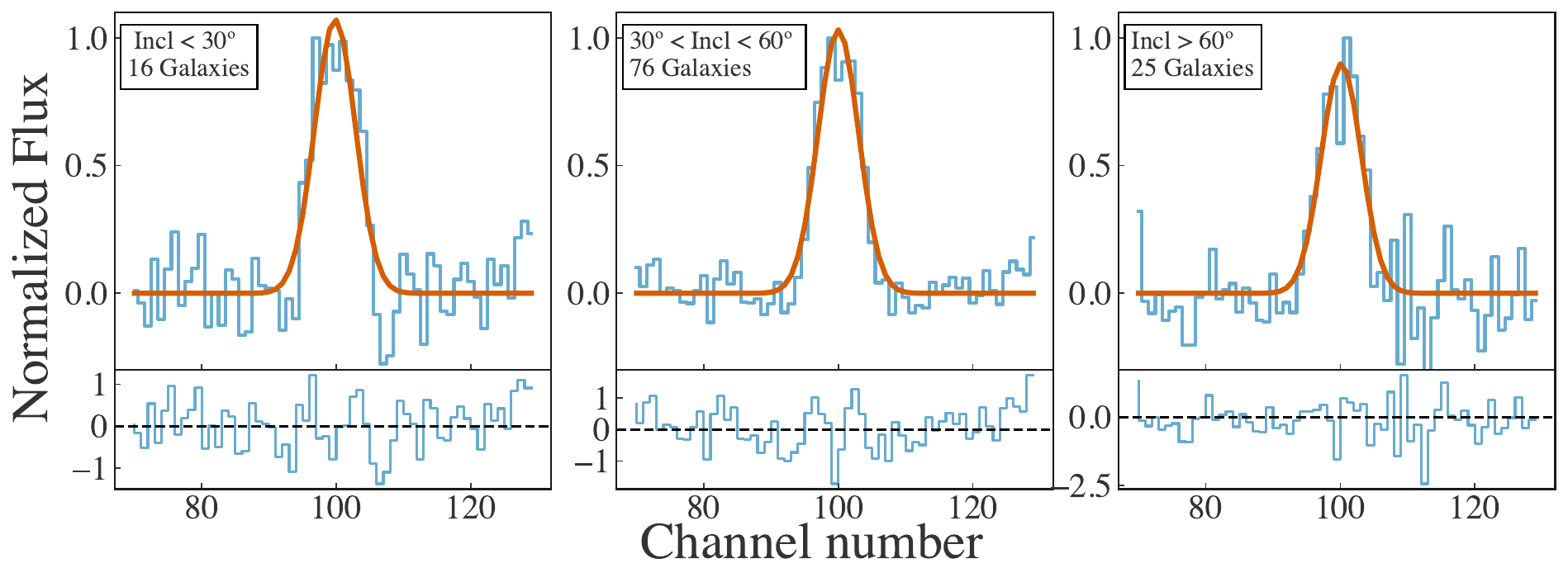}
    \caption{\textit{Upper Panel}: Stacked spectra (light blue curves) for the three bins in inclination. Similar to Fig. \ref{fig:full_stack}, we plot the best-fit single Gaussian model to the profiles (solid orange) and the fit residuals normalized by the error below each plot. As in Fig. \ref{fig:full_stack}, there is no broad component signature, i.e., no outflow detection in any of the stacks.}
     \label{fig:stack_inc}
\end{figure*}

\begin{figure*}
    \centering
    \includegraphics[width=0.90\linewidth, height=0.90\textheight]{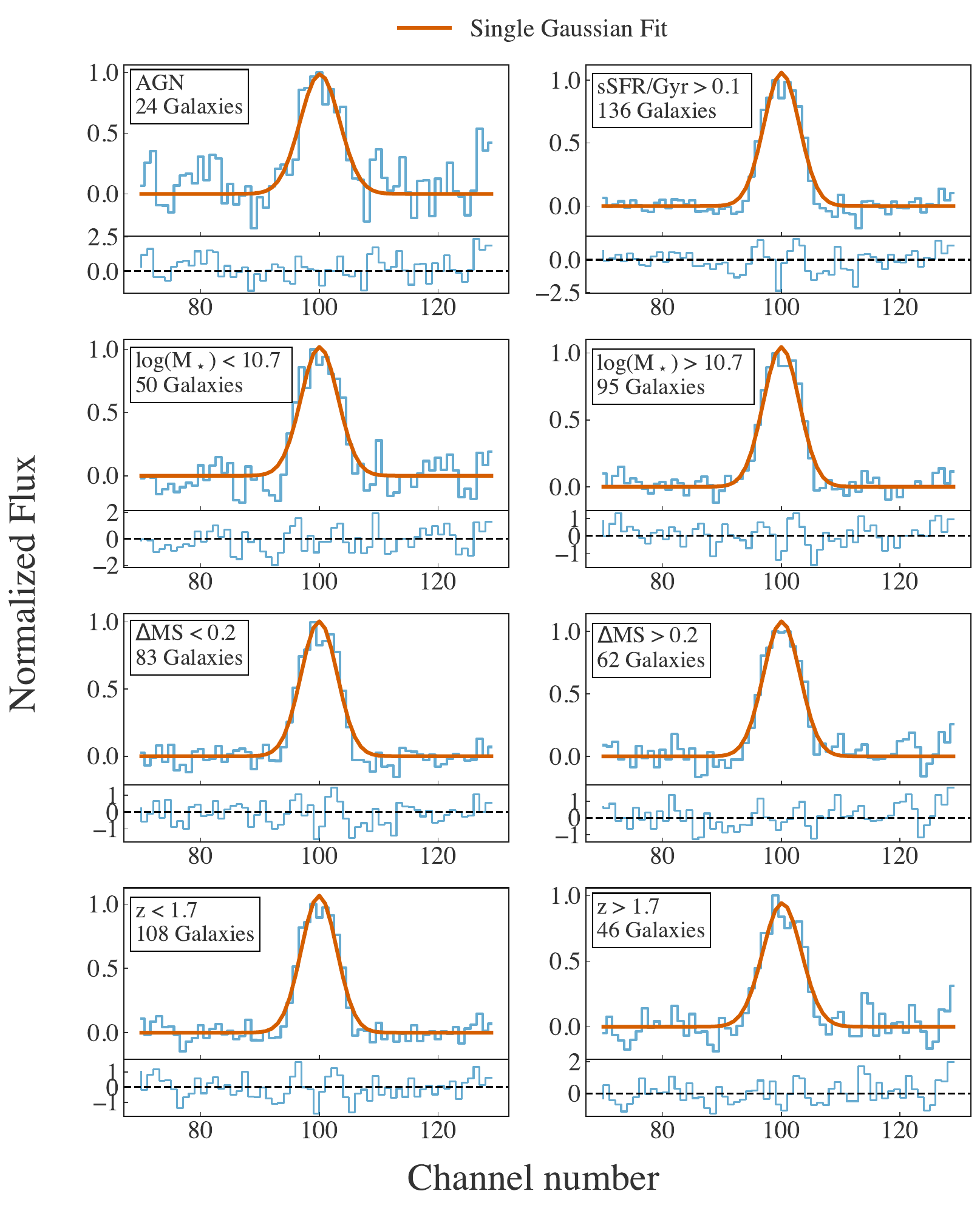}
    \caption{ Stacked spectra (blue curves) for all the 8 subsamples described in Sec. \ref{sec:stack}. Overlaid on the spectrum is the best-fit single Gaussian profile (solid orange). Fitting shows that no significant outflow component is detected in the stacked spectrum. {\textit{Bottom Panel}:} Similar to Figs \ref{fig:full_stack} and \ref{fig:stack_inc}, normalized fit residuals ($\mathrm{\frac{data-model}{noise}}$) from the single Gaussian fitting.}
    \label{fig:stacks}
\end{figure*}

The results from the stacking of the spectra are shown in Fig. \ref{fig:full_stack} for the full sample and in Figs. \ref{fig:stack_inc} \& \ref{fig:stacks} for subsamples  defined in Sec. \ref{sec:stack}. The stacked lines are fit both with a single and double Gaussian component, where the fit has the following constraints: (1) one component cannot have both the larger amplitude and the larger width, (2) the amplitude of both components should be positive, (3) the center of the second component should be within 10 channels of the center of the first component, to avoid fitting noise peaks on the edges of the spectrum. In addition, for each channel, we define as the error on the flux the error derived in Sec. \ref{sec:stack} from the stacked ``blank" spectra, scaled by the number of sources stacked in that particular channel ($\sigma_{\mathrm{noise}} \propto 1/\sqrt{N_{\mathrm{obj}}}$ where $N_{\mathrm{obj}}$ is the number of sources stacked). This number varies from channel to channel as all spectra do not have the same number of spectral channels (even prior to re-binning). When aligning them on the emission line and averaging channel-per-channel, the edges, in particular, will be the average of fewer channels.

In all cases, the single Gaussian fit residuals, normalized by the noise in the spectra, are displayed below the stacked spectrum in Figs. \ref{fig:full_stack}, \ref{fig:stack_inc} \& \ref{fig:stacks}. Since some of the individual integrated galaxy spectra display the double-peaked profile expected from inclined rotating disks, this feature sometimes propagates in the stacked spectra and can be seen in the residuals. However, no clear outflow signature is visible.

For completeness, we perform a two-Gaussian fit to the spectra and observe that it does not improve the goodness-of-fit. Indeed, the reduced $\chi^2$ of the single Gaussian fit is systematically lower than that of the double Gaussian fit. We also perform an analysis of the Bayesian information criterion (BIC, \citealt{schwarz1978})\footnote{The BIC is defined as $\mathrm{BIC} = \mathrm{kln(n)} + \chi^2$, where k is the number of free parameters, and n is the number of fitted data points \citep{hogg2010}.} of our single and double Gaussian fits. To do so, we define $\Delta\mathrm{BIC} = \mathrm{BIC_{sing}} - \mathrm{BIC_{doub}}$ as the difference between the BIC of the single Gaussian fit and the double Gaussian fit, respectively, where $\Delta\mathrm{BIC} > 0$ favors the double Gaussian model to fit the data, and $\Delta\mathrm{BIC} > 10$ is strong evidence that a double Gaussian fit is needed to fit the data \citep{liddle2007}. For all stacked spectra, $\Delta\mathrm{BIC} < 0$, in line with the result that there are no outflow signatures in the stacks. In addition, the flux of the very low amplitude broad component is systematically below the noise level (see Fig. \ref{fig:flux_lim}, as determined following the procedure outlined in Sect. \ref{sec:flux_lim}). Therefore, there is no statistically significant evidence for a broad component associated with outflows in the stacked spectra.

\subsection{Investigating the Extended Emission}\label{sec:big_ap}

Since outflows in various phases have been observed in extended regions in the outskirts of their host galaxies (e.g., \citealt{genzel2011, genzel2014, fabian2012, newman2012a, newman2012b, fs2014, fs19, leroy2015, hc19, rebecca2020}), we also looked into the extended emission around our sample. To do so, we investigate the spectra extracted from both larger apertures and annuli apertures centered on the galaxies. The larger apertures are designed by ``dilating" the original mask by 10kpc in every direction, while the ``annuli" apertures are the difference between the bigger and original apertures. The 10kpc increase in size is chosen based on the extent of the molecular gas outflow in \cite{hc19}. Finally, we perform the same stacking analysis and fitting as previously on these newly extracted spectra. While both methods do reveal spatially extended emission that is undetected in individual cubes and thus missed by the nominal masks employed, fitting the double Gaussian profile to the lines returns the same result as for the initial aperture: we observe no faint broad component in the stacked profile, and the fit residuals show no improvement from performing a two-component Gaussian fit (see Appendix \ref{ap:ap} for the stacked profiles from the extended and annuli apertures respectively).

\subsection{Subset-sampling Analysis}\label{sec:subsamp}
We created the sub-samples described in Sec. \ref{sec:stack} based on observed correlations between ionized gas outflow incidence and galaxy properties. However, if molecular gas outflows have different correlations or are less prevalent in galaxies, a non-matched choice of parameter space leaves too much ``dilution" from sources with weak or absent outflow signatures. We apply the subset-sampling analysis described in \cite{stanley2019} to search for potential outflows in sub-samples that may not match the subsets defined by M$_\star$, $\Delta$MS, or AGN as defined in Sect. \ref{sec:stack} based on ionized gas outflow studies.

Following the methodology, we design a subset of spectra by randomly selecting between 3 and 154 galaxies from the full sample. These are then stacked and fitted with our two-component Gaussian model. Each spectrum in the subset is assigned a grade based on the strength of the broad Gaussian fit component (i.e. how much flux is contained in the secondary broad component, where the flux is defined as $F_\mathrm{output} = \sqrt{2 \pi} \times A_\mathrm{out} \times \sigma_\mathrm{out}$, with $A_\mathrm{out}$ the best fit broad component amplitude and $\sigma_\mathrm{out}$ is the best fit broad component width). The process is repeated 100,000 times, each time randomly selecting spectra from the full sample, such that every spectrum is graded. This allows us to select a final sub-sample of galaxies with the highest grade, i.e., the best candidates to have an outflow. We then repeat the stacking process on the galaxies with the highest grades and evaluate if we retrieve any underlying broad component. \par
\begin{figure}
    \centering
    \includegraphics[width=\linewidth]{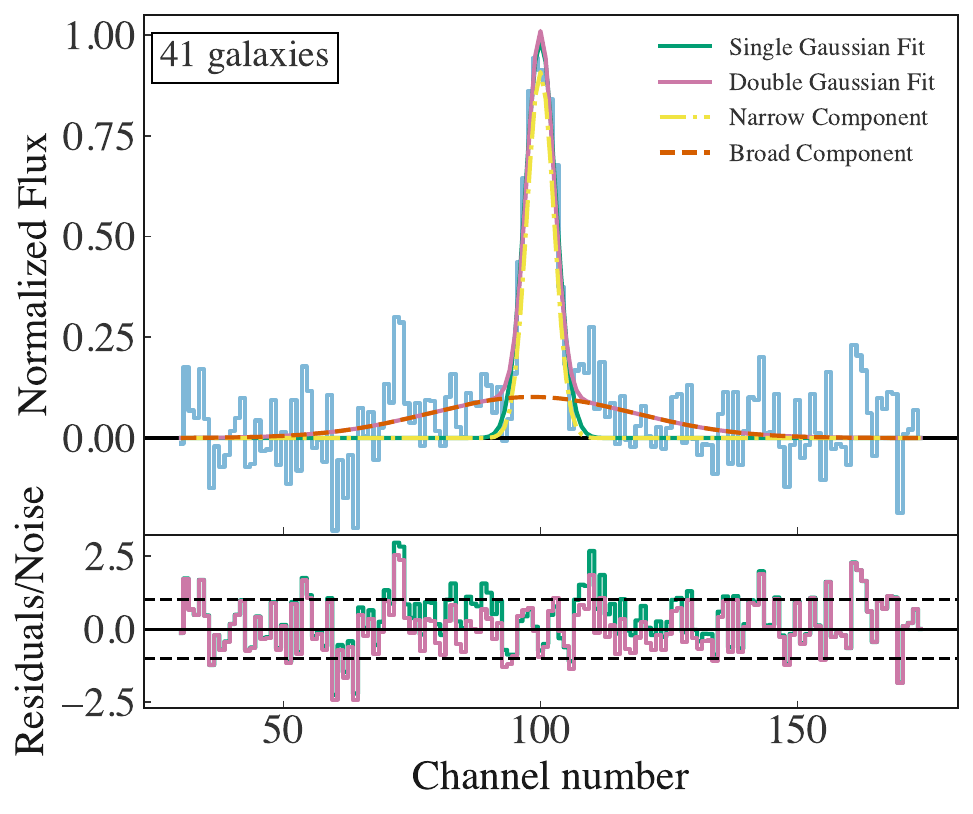}
    \caption{Stacked spectrum of the 41 galaxies with the highest grade in the subset sampling analysis, along with the best-fit single Gaussian (solid Green), double Gaussian (solid purple). The two individual components of the double Gaussian fit are also plotted (dot and dashed yellow for the narrow component and dashed orange for the broad component). The bottom panel shows both models' best-fit residuals (normalized by the noise). The black dashed lines indicate where $\mathrm{\frac{data-model}{noise}} = 1$.}
    \label{fig:subsamp}
\end{figure}

Using this method, we find a subset of 41 galaxies (26\% of the full sample, marked with a $\dagger$ symbol in Table \ref{tab:sample}) for which the stacked spectrum displays a tentative outflow signature, evaluated at 4$\sigma$, shown in Fig. \ref{fig:subsamp}. The residuals of the best-fit show improvement when using a double Gaussian model, and its reduced $\chi^2$ is closer to one than than that of the single Gaussian fit. Repeating the analysis of the BIC presented in Sec. \ref{sec:fit}, we find $\Delta\mathrm{BIC = 6}$, which favors the double Gaussian fit as better for the spectrum. All but one galaxy have coverage by either \textit{Chandra}, MOSDEF, \textit{Spitzer} or VLA, and 13 are flagged as AGN (Sec. \ref{sec:agn}). We investigate the subsample properties in Appendix \ref{ap:subsamp} and find no striking correlation between the tentative outflow detection and the galaxies' physical properties, except for a slight bias towards higher masses ($\langle$log(M$_\star$/M$_\odot$)$\rangle$ = 10.95). This might be consistent with the trend in M$_\star$ reported by, e.g., \citealt{fs19} in the case of predominant AGN-driven winds. We also investigate the outflow properties (outflow mass, mass outflow rate, mass loading factor) from the detection in Sec. \ref{sec:flux_lim} and report the values in Table \ref{tab:avg_prop}. Still, the low significance of the result and lack of outflow signature in the stacks of high-mass and AGN-hosting galaxies prevent any firm conclusion with the data in hand. In addition, a caveat of the subsampling analysis is that it can result in selecting the galaxies whose noise properties will stack positively, mimicking an outflow detection. We try varying the re-binning of the spectra and stacking the spectra extracted from the extended apertures of the galaxies and observe that the feature remains. 


\subsection{Upper Flux Limit}\label{sec:flux_lim}

\begin{figure*}
    \centering
    \includegraphics[width=\linewidth]{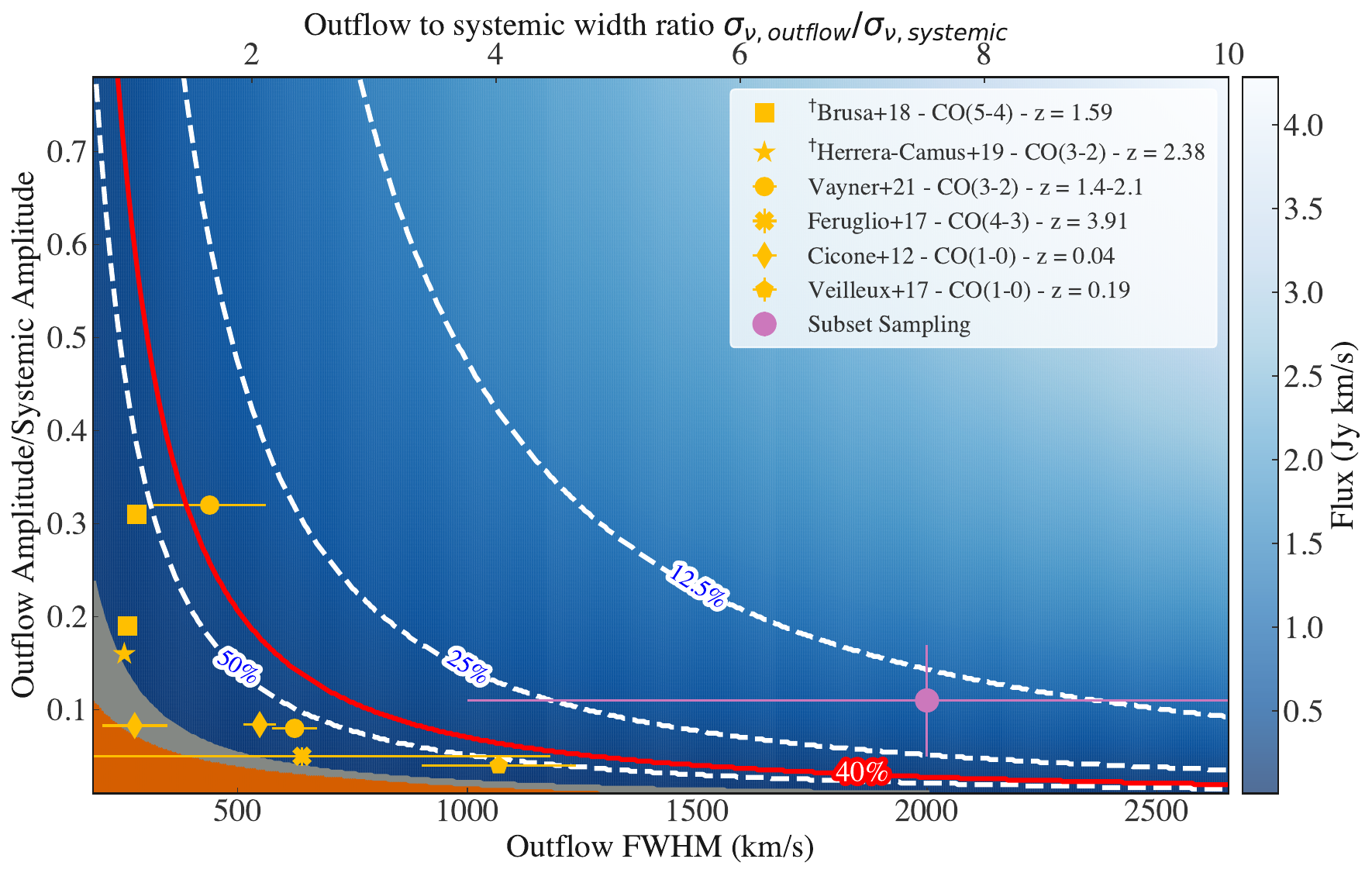}
    \caption{Parameter space of outflows based on their properties (amplitude and FWHM) with respect to the systemic emission. The background indicates the flux for a given set of outflow properties, where lighter indicates higher flux. The grey area delimits the parameter space spanned by outflows below the noise of our stacked spectrum if all sources in the sample have outflows, whereas the orange shaded area shows the flux within the best fit ``broad" Gaussian component of the full sample stack, which stands below the noise level. The purple circle indicates the results from the tentative outflow detection in the subset of 41 galaxies with the highest grades. The orange markers indicate outflow properties of molecular outflows identified in bright AGNs/quasars found in the literature \citep{cicone12, feruglio2017, veilleux17, brusa2018, hc19, vayner2021}. The $\dag$ symbol marks values reported in the literature but not quoted precisely and/or without errors. Based on the upper flux limit from our stacking analysis, the white dashed and the red lines represent a lower limit for detection in the amplitude ratio vs. outflow FWHM parameter space. The incidence for the sample based on demographics of ionized gas outflows \citep{fs19}, 39.5\%, is indicated with the solid bright red line.}
    \label{fig:flux_lim}
\end{figure*}

From our non-detections, we measure an upper limit on the flux an outflow could have based on the noise level measured in the empty stacked spectrum. Defining a detection above the noise level as 3$\sigma_\mathrm{noise}$, where $\sigma_\mathrm{noise}$ is the standard deviation per channel of our empty spectrum (see Sec. \ref{sec:stack}), we compute the upper flux limit of a putative broad component using:

$$ F_{UL} = 3 \sigma_\mathrm{noise} \sqrt{N_\mathrm{ch}}$$

where $N_\mathrm{ch} = 2 \times 3 \times \sigma_\mathrm{out, ch}$ is the number of channels spanned by this hypothetical Gaussian, defined as 3 standard deviations $\sigma_\mathrm{out, ch}$ (in channels) on both sides of the Gaussian center to encompass $> 99$ \% of the flux\footnote{This is a conservative approach; if considering the flux within the FWHM, the upper limits would be lower by a factor of 1.6}. As we do not have a value of $\sigma_\mathrm{out}$, we compute this upper limit for values of $\sigma_\mathrm{out}$ between $0.07 - 10 \times$ the narrow component width $\sigma_{\nu, \mathrm{systemic}}$ (see Fig. \ref{fig:flux_lim}).

To investigate what this limit represents for the outflow and what parameter space this hypothetical outflow might span, we create a grid of 500 flux values for a set of broad-to-narrow amplitude ratio ($0.01 - 0.8$) and broad-to-narrow linewidth ($0.07 - 10$; which is also translated to physical values using the average channel width and average line amplitude of the stacked spectra). We plot this grid as the background of Fig. \ref{fig:flux_lim}, as function of the input parameters. We note that for outflow width below $2 \times$ the galaxy line width, it will be almost impossible to separate the galaxy emission from the outflow emission, if the lines have the same central velocity (see also Sec. \ref{sec:test}, Fig. \ref{fig:simu}). On the other hand, if the outflow central velocity is shifted with respect to the galaxy emission line's, it should be possible to detect an outflow with a small width.

This upper flux limit is computed by assuming the maximum possible flux value a stacked outflow component might have under the noise level. If all the galaxies in our sample have an outflow participating in the average - i.e. if the outflow incidence is 100\% -, the flux limit computed above corresponds to the grey shaded area in Fig. \ref{fig:flux_lim}. However, previous studies (e.g., \citealt{fs19}) report lower incidence values, which also depend on stellar mass, redshift, and SFR. In this case, the stacking of spectra with and without outflow components will ``average out" the strength of the potential outflows, decreasing the stacked flux. Consequently, the corresponding maximum flux of outflows in the sample increases as the incidence of outflows decreases. We show this effect by overlaying on the plot upper flux limits computed for incidences of 50\%, 25\% and 12.5\% (dashed white lines). This increases the flux limit as it would require the stacked outflow to have 2$\times$, 4$\times$, and 8$\times$, respectively, the flux of the outflow computed for 100\% incidence. Finally, using the ionized gas outflow incidence values from \cite{fs19}, we compute the overall incidence of the sample by taking into account the stellar mass and distance from the main sequence of the galaxies in the sample. This corresponds to an overall 39.5\% outflow incidence in the sample of 154 galaxies, plotted in Fig. \ref{fig:flux_lim} as the solid red line. In addition, we add to the plot the results from the subset sampling analysis (purple circular marker), where the stack of 41 galaxies reveals a tentative outflow signature. Compared to the full sample, this corresponds to an outflow incidence of 26\%, which is in marginal agreement with the upper limits derived from the noise.

For comparison with other molecular gas outflows studies, we also overlay on Fig. \ref{fig:flux_lim} the positions of cosmic noon detections of molecular gas outflows in individual galaxies by \cite{feruglio2017, brusa2018, hc19} and \cite{vayner2021}, in addition to two local Universe individual detections reported in \cite{cicone12} and \cite{veilleux17}. All but the lowest redshift detection in Mrk 231 by \cite{cicone12} lie above the flux limit for 100\% outflow incidence. This could be a result of the proximity of the target which allows the detection of weaker molecular gas outflows. Unfortunately, very few CO detections of outflows exist at cosmic noon, and all originate from luminous AGNs or QSOs (except for the one reported in \citealt{hc19}), which are not representative of the main galaxy population (see \citealt{veilleux2020} for a review).


\section{Tests of the Method}\label{sec:test}
To check how the different steps of our analysis (see Sec. \ref{sec:methods}) might influence the results, we redo the analysis on mock data with outflow signatures. This exercise also allows us to quantify limits on the strength and width of a broad outflow component. In addition, we also investigate the potential effects of re-binning on the retrieval of outflow signatures in the stacked spectra.

\subsection{Recovery Analysis with Mock Data}\label{sec:mock}
\subsubsection{Creating Mock Outflows}
For this exercise, we create a mock spectrum for individual galaxies employing the single Gaussian best fit to the data, adding a broader component centered on the narrow component and peak-normalizing the combined narrow+broad (outflow+galaxy) spectrum. Using the blank spectrum extracted for each galaxy to evaluate the noise in each spectrum, we inject realistic (normalized) noise in the mock outflow+galaxy line profile and thus recover a mock galaxy emission line with an outflow component and a SNR comparable to the original data. This way, we have 154 mock galaxy spectra with an outflow signature and realistic noise, which we re-bin according to the method described in Sec. \ref{sec:re-binning}, and then stack following the method described in Sec. \ref{sec:stack}. Finally, we fit the stack with the same procedure described in Sec. \ref{sec:fit}. We repeat the process 10,000 times, varying the input outflow amplitude between $0.005 - 0.8 \times$ the narrow component amplitude and the input outflow width between $1 - 5 \times$ the galaxy line width.

For the purpose of this analysis, we choose to add the broad outflow component as centered on the narrow component. That might not be the case, as outflows often show a velocity shift with respect to the narrow emission line. However, when averaging spectra with outflow signatures in the stacking process, the resulting average spectra will display outflow signatures as a broad component centered on the narrow component, hence, we chose to add the broad component already aligned with the narrow component. 



\subsubsection{Output Flux Recovery}
\begin{figure*}
    \centering
    \includegraphics[width=\linewidth]{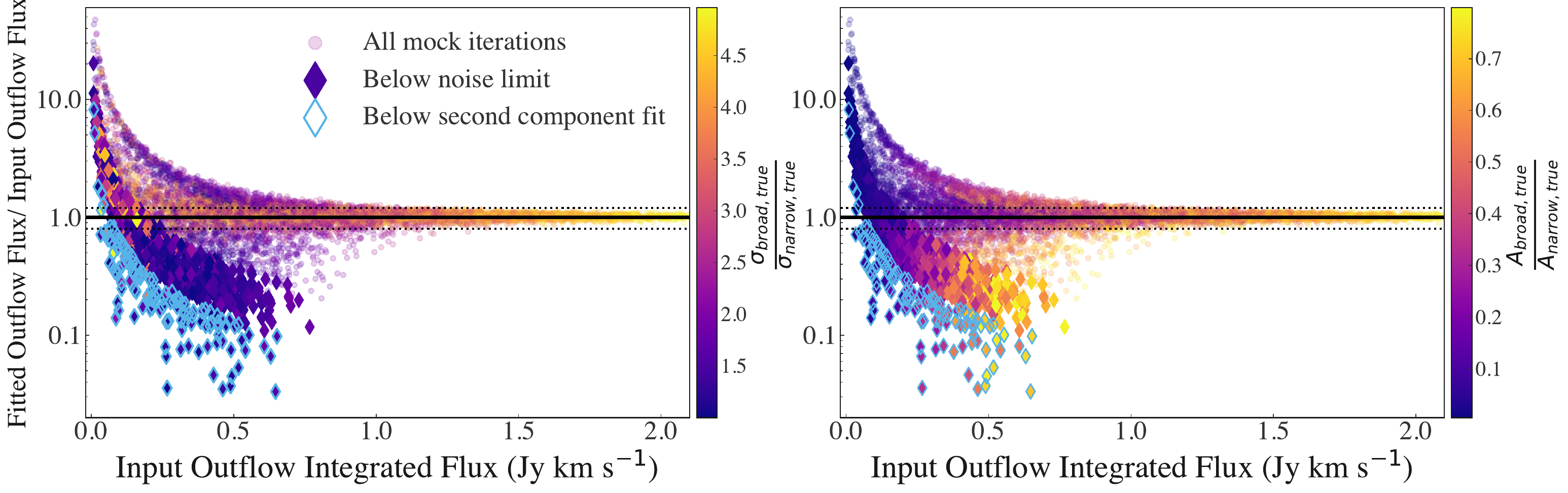}
    \caption{Recovered outflow flux from stacking vs. outflow flux computed from the input parameters for the 10,000 mock stacked spectra. The x-axis is converted from normalized flux values to real units using the average amplitudes and velocity binsizes of the stacked spectra. For strong input outflows (high broad-to-narrow amplitude and/or width), the fitting of the stacked spectra shows excellent recovery of the input flux, which supports that the non-detection of outflows in our subsamples is not a product of the method. \textit{Left Panel:} Color-coding by the input broad-to-narrow width ratio. \textit{Right Panel:} Same as the left panel, but this time looking at the distribution of the broad-to-narrow amplitude ratio. The solid black line in both panels shows perfect recovery, whereas the two dashed lines mark the area within 20\% of this ``perfect" fit. The solid diamonds show the iterations where the fitted outflow flux is below the noise of the stacked spectrum. The blue diamond contours show the iterations where the fitted mock outflow flux is below the flux of the best fit ``broad" Gaussian component of the full sample stack.}
    \label{fig:simu}
\end{figure*}

The results are plotted in Fig. \ref{fig:simu}, which shows, as a function of the input parameters, how well the outflow fluxes are recovered. To make the comparison between the input mock outflows and the retrieved mock outflows, we define the input outflow flux as $F_\mathrm{input} = \sqrt{2 \pi} \times \overline{A_\mathrm{in}} \times \overline{\sigma_\mathrm{in}}$ where $\overline{A_\mathrm{in}} $ is the mean input outflow amplitude over the whole sample and $\overline{\sigma_\mathrm{in}}$ is the mean input outflow width over the whole sample, and define the output flux as $F_\mathrm{output} = \sqrt{2 \pi} \times A_\mathrm{out} \times \sigma_\mathrm{out}$ where $A_\mathrm{out}$ is the fitted outflow amplitude after stacking, and $\sigma_\mathrm{out}$ is the fitted outflow width after stacking. From Fig. \ref{fig:simu}, we observe a clear trend between the strength of the outflow and how well the fitting retrieves the input outflow, as expected. In addition, the width of the outflow seems to have a greater impact on how well the fit performs compared to the amplitude, except for broad-to-narrow amplitude ratios below $0.1$. Finally, the excellent recovery (within 20\% or better) for sufficiently high flux, broad component amplitude, and $\sigma$ supports that the re-binning procedure described in Sec. \ref{sec:re-binning}, which we also apply to all mock spectra before stacking, does not influence the ability to recover a sufficiently strong and wide outflow component.

We note that the fanning up of the distribution at low fluxes is an artifact of the fitting reflecting the difficulty in constraining two components at low fluxes, hence low S/N ratios.  Examination of the results shows that in those cases, the formal ``best-fit" model is the sum of two Gaussians of very similar amplitude and width (one component with slightly lower amplitude and slightly higher width and \textit{vice versa}, as imposed by the constraints reported in Sec. \ref{sec:fit}). This returns high flux values for the ``broad" component, which translates, when computing the ratio of the recovered flux to the input flux, to values much higher than 1.

\subsubsection{Comparison with the Real Data}
Finally, we compare the analysis of the mock outflows to that of our stacked spectrum from Fig. \ref{fig:full_stack} to determine an upper limit on the broad outflow component. First, we consider our best-fit double Gaussian for the stack of the full spectra. This model includes a best-fit broad component, which is negligible in all cases (as investigated in Sec. \ref{sec:fit}). To investigate if this scenario is reproducible with the mock outflows, and under what circumstances (i.e. for which outflow properties), we estimate the flux within this negligible ``broad" component and identify which mock input outflow properties would reproduce the result we observe with the real data, or return lower flux values (blue diamonds contours in Fig. \ref{fig:simu}). Similarly, we use the noise of our stacked spectra as an upper limit on the flux of a potential undetected outflow and repeat the same exercise as for the second component (solid diamonds in Fig. \ref{fig:simu}). We conclude that for fluxes below these limits: 1) for the most part, the fit does not recover the ``true" input flux, and 2) those flux values correspond either to low values of flux or to outflows with low velocities ($< 1.5 \times$ the width of the narrow component).

\subsection{Effect of Re-binning on the Stacked Spectra} \label{sec:nore-bin}
\begin{figure}
    \centering
    \includegraphics[width=\linewidth]{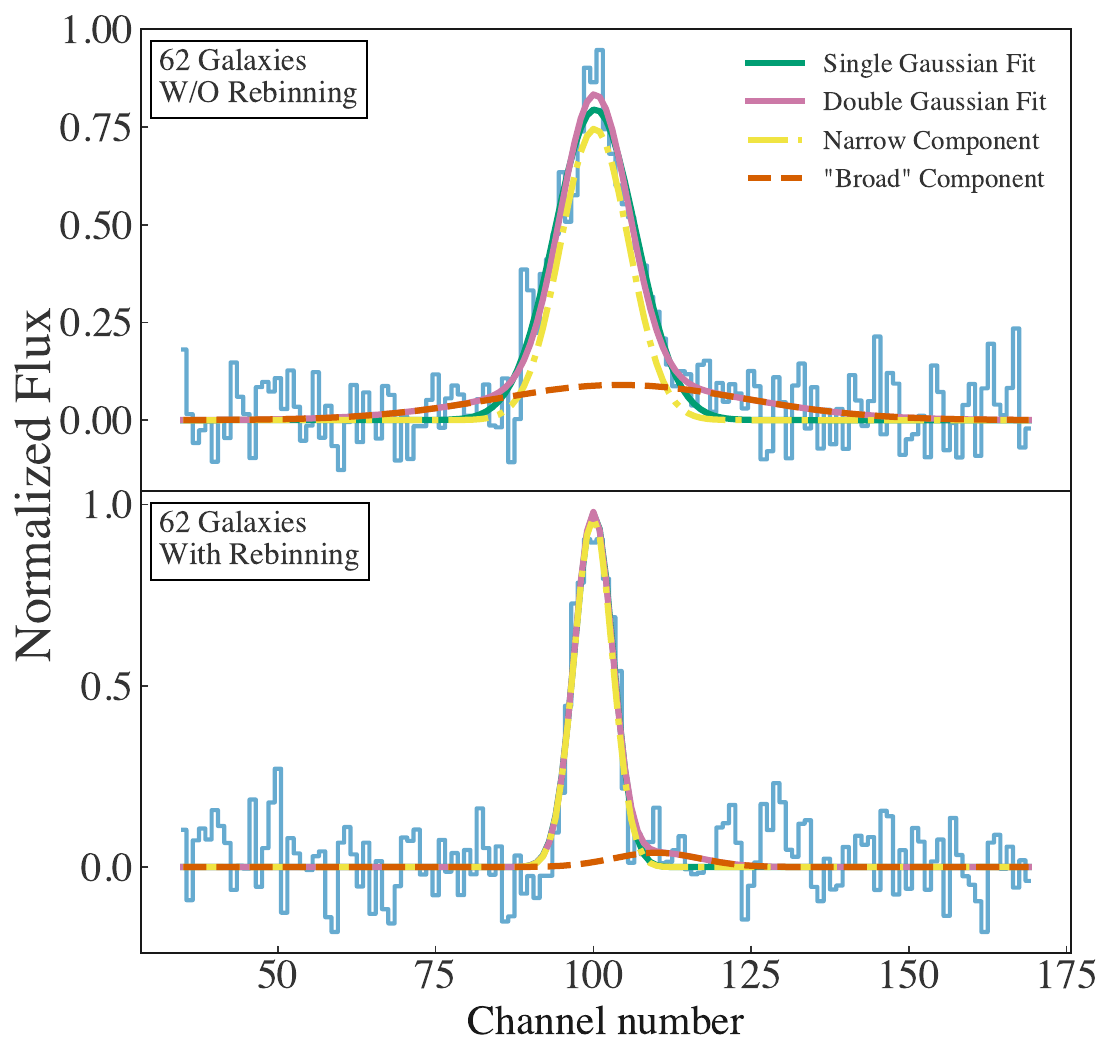} 
    \caption{Example plot of the stacking of galaxy spectra without (upper panel) and with (lower) re-binning the individual spectra according to their width pre-stacking. The stacked spectra include the 62 galaxies with $\Delta$MS $> 0.2$, where we overlay the best fits to the line: solid green is the single component Gaussian, solid purple is the double component Gaussian, and the dashed yellow and orange show the two separate components of the double Gaussian fit. Without re-binning the velocity channels, the stacking induces an artificial broad component mimicking an outflow signature (top panel), which disappears when including the re-binning (bottom panel).}
    \label{fig:nore-bin}
\end{figure}

A possible caveat of the method is re-binning before stacking the spectra. As mentioned in Sec. \ref{sec:re-binning}, stacking line profiles with a wide range of FWHM can result in the creation of artificial broad wings, which can be misinterpreted as an outflow component. This effect is illustrated in Fig. \ref{fig:nore-bin} for the stack of 62 $\Delta$MS $> 0.2$ galaxies, where we stack the spectra before and after re-binning to compare the effect of the varying linewidths of the emission lines on the stacked profile. On the other hand, re-binning may reduce the ability to detect a broad component through dilution.

To investigate how re-binning affects the retrieval of outflows in our sample we try varying the values we re-bin the spectra to, which all return similar results to those displayed in Figs. \ref{fig:full_stack} and \ref{fig:stacks} and where $\Delta\mathrm{BIC} < 0$ favoring a single-Gaussian component fit in all cases. In addition, we also try stacking together subsamples of galaxies with similar linewidths (in which case there is no need for re-binning) and observe no outflow component once again. Since the integrated linewidth can also serve as a proxy for galaxy inclination (for galaxies of the same mass, i.e. rotating at the same velocity, the more inclined galaxies' line profiles will be wider), the stacking of galaxies with similar linewidths without rebinning also ensures that we do not miss outflows based on inclination.

Finally, the procedure described in Sec. \ref{sec:test} in which we produce mock outflow emission, shows that re-binning does not prevent the fitting from detecting a broad secondary component when a clear signature is present. As a result, if the PHIBSS sample had broad outflow signatures present, which could be revealed by stacking, the re-binning process would not hinder our ability to do so.

\section{Discussion}\label{sec:disc}

The work presented here investigates the presence of outflows in a sample of 154 SFGs at $0.5 \leq z \leq 2.6$ in a sample representative of the general population. Based on the results presented in Sec. \ref{sec:res}, we consider here the physical limits on molecular gas outflows we can retrieve from our stacked spectra, as well as discuss the contribution of various gas phases to outflows.

\subsection{Upper limit on Molecular Gas Mass and Mass Outflow Rate}
We derive upper limits on outflow properties such as the mass outflow rate and the mass loading factor, for all stacked samples. For the samples for which we do not have an outflow detection, we use as outflow widths and velocities the average values from ionized gas outflows reported in \cite{fs19}, extracted from their stacked spectra. We note that molecular gas outflows traced by CO are generally reported to be slower than their ionized gas counterparts, and thus, these computed values should be interpreted as very conservative upper limits. For the stacked spectrum from the subset sampling analysis in which we have a detection above the noise level (see Sec. \ref{sec:subsamp}), we define the outflow velocity as $\mathrm{v}_\mathrm{out} = |\Delta \mathrm{v} - 2 \sigma_\mathrm{out}|$, where $\Delta \mathrm{v}$ is the central velocity differential between the narrow and broad component, and $\sigma_\mathrm{out}$ is the width of the best-fit broad component \citep{veilleux2005, genzel2011, genzel2014, concas2022}. All values are reported in Table \ref{tab:avg_prop}.

To translate the relative limits discussed in the previous subsection into physical limits, we compute the CO line luminosity $L'_{CO}$ in K km s$^{-1}$ pc$^2$ using the following relation \citep{solomon1997, veilleux17, tacconi2020}:

\begin{equation}
    L'_{CO} = 3.25 \times 10^7 \cdot S_{CO} \Delta V \cdot \frac{D_\mathrm{L}^2}{\nu^2_\mathrm{obs} (1 + z )^3},
\end{equation}

where the luminosity distance $D_\mathrm{L}$ is in Mpc, the observed frequency $\nu_\mathrm{obs}$ is in GHz, and the integrated line flux $S_{CO} \Delta V$ is in Jy km s$^{-1}$. We use the median values of the galaxies that go into the stacking for the redshift, observed frequency, and luminosity distance.

This luminosity can then be converted into an outflow molecular gas mass estimate:

\begin{equation}
    M_\mathrm{out, mol} = \alpha_{CO} \cdot \frac{L'_{CO}}{r_{31}}
\end{equation}

where $r_{31} = 0.77$ is the ratio of temperature brightness to correct for the fact that we are not observing the CO(1-0) transition line but the CO(3-2) \citep{boogaard2020}, and $\alpha_{CO} = 0.8$ is the ULIRG-like H2-CO conversion factor in units of M$_\odot$ (K km s$^{-1}$ pc$^2$)$^{-1}$ \citep{veilleux17, fluetsch19}. We use the ULIRG $\alpha_{CO}$ in this study, which is commonly used in the literature to obtain the outflow molecular gas mass, but this value relies on the assumption that outflows are well mixed and high metallicity, which might not be the case. Other commonly invoked values of $\alpha_{CO}$ range from 0.3 for the optically thin case (e.g., \citealt{bolatto2013, richings2018}) to 4.3 for the Milky Way value \citep{bolatto2013b}. Studies in the Local Universe indicate that $\alpha_{CO}$ in outflows is variable from case to case, however the lack of consensus and the resolution of our data does not allow us to take this into account (see \citealt{veilleux2020} for a recent review). Since the molecular gas mass estimate is directly proportional to $\alpha_{CO}$, this represents one of our estimates' major sources of uncertainty. 

Using the estimated molecular gas mass, we can compute an upper limit for the outflowing mass rate, using a simplistic yet plausible assumption commonly used in the literature (e.g., \citealt{sturm2011, fiore2017, hc19}; with the details of the derivation in \citealt{rupke2005}):

\begin{equation}
    \dot{M}_\mathrm{out, mol} = M_\mathrm{out, mol} \times \frac{\mathrm{v}_\mathrm{out}}{R_\mathrm{out}}
\end{equation}

Here, $R_\mathrm{out}$ is the outermost radius reached by the outflow (we adopt the median effective radius of the galaxies stacked derived from \textit{HST} imaging of the data \citep{tacconi2013}, by lack of further spatial information and following what was done in \citealt{newman2012a} and \citealt{fs19}) and $\mathrm{v}_\mathrm{out}$ is the outflow velocity. We compute $\dot{M}_\mathrm{out, mol}$ for each stacked sub-sample.

Using this mass outflow rate, we compute the mass loading factor $\eta = \dot{M}_\mathrm{out} / $SFR, where SFR is the median SFR of the sample. The resulting values for each sub-sample are in the ranges $\dot{M}_\mathrm{out, mol} = 380 - 2300 \: \mathrm{M_\odot/yr}$ and $\eta_\mathrm{UL} = 2.2 - 35.4$. All values and average galaxy properties used for these estimates are reported in Table \ref{tab:avg_prop}. In all cases, the derived upper limits on the mass loading factor allow for molecular gas outflows to carry a substantial amount of mass, enough to dominate the gas depletion in the galaxy. However, these values rely on numerous assumptions made to yield very conservative upper limits. 

These limits are in line with some results from state-of-the-art simulations, which find that the cool ($< 10^4$K) gas dominates the mass budget of AGN- and SF-driven outflows, with mass loading factors $\eta \sim 1-10$ \citep{rathjen2023, ward2024}. In more extreme cases, \citealt{kim2020} and \citealt{porter2024} find $\eta \sim 100$, which appears to be ruled out by our upper limits. However, the mass loading factors in \citealt{kim2020} are measured at one scale height above the mid-plane, and the majority of the winds have very low velocities ($\sim 10 - 100$ km/s). As a consequence, the mass loading factors at larger distances are much lower and in better agreement with our upper limits.

However, in most cases, the simulations cannot resolve gas colder than $10^4$ K, limiting the conclusions that can be drawn on the dominant gas phase. Comparing with theoretical predictions, ``bathtub models" usually suggest an $\eta_{out}$ of unity \citep{dekel2013}. In addition, \cite{dekel2014} estimate that to match a simplistic bathtub toy model to observations, the \textit{net} mass loading factor (the mass loading factor of the outflow minus the mass loading factor of recycled gas falling back in the galaxy) at $z \sim 2$ should be 0, i.e. none of the outflowing gas escapes the galaxy, and all is recycled. This result can be in line with our upper limits on $\eta$ as we only take into account the outflowing gas mass loading factor $\eta_{out}$.



\begin{deluxetable*}{c|ccccc|ccc}
    \tablehead{
    \colhead{Sample} & \colhead{$\langle z\rangle$} & \colhead{$\langle$log(M$_\star$/M$_\odot$)$\rangle$} & \colhead{$\langle$ R$_{e} \rangle$} & \colhead{$\langle \nu_\mathrm{obs} \rangle$} & \colhead{$\langle$ SFR $\rangle$} & \colhead{v$_\mathrm{out}$} & \colhead{$\dot{M}_\mathrm{out, mol}$} & \colhead{$\eta_{\mathrm{UL}}$} \\ 
    \colhead{} & \colhead{} &  \colhead{} & \colhead{kpc} & \colhead{GHz} & \colhead{M$_\odot$/yr} & \colhead{km/s} & \colhead{M$_\odot$/yr} & \colhead{}
    }
    \centering
\startdata
        All & 1.05 & 10.8 & 4.35 & 136.9 & 50.1 & 450 -- 2300$^\dagger$ & 135 -- 1567  & 2.7 -- 31.3 \\
        AGN & 1.09 & 11.0 & 3.25 & 140.6 & 112.9 & 1360$^\dagger$ & 1619 & 14.3 \\
        $M_* < 10^{10.7} \: M_\odot$ & 1.01 & 10.5 & 4.30 & 134.2 & 31.6 & 380$^\dagger$ & 78 & 2.5\\
        $M_* > 10^{10.7} \: M_\odot$ & 1.01 & 11.0 & 4.45 & 140.7 & 63.1 & 420 -- 1360$^\dagger$ & 237 -- 1383 & 3.8 -- 21.9 \\
        $\Delta$MS $> 0.2$ & 1.13 & 10.7 & 3.60 & 137.7 & 79.4 & 390 -- 1500$^\dagger$ & 373 -- 2814 &  4.7 -- 35.4 \\
        $\Delta$MS $< 0.2$ & 1.02 & 10.9 & 5.00 & 137.1 & 31.6 & 450 -- 1520$^\dagger$ & 85 -- 525 & 2.7 -- 16.6 \\
        z $> 1.7$ & 2.21 & 10.8 & 3.50 & 158.5 & 108.1 & 460 -- 1000$^\dagger$ & 345 -- 1105 & 2.2 -- 6.9 \\
        z $< 1.7$ & 0.76 & 10.8 & 5.25 & 144.0 & 31.6 & 460 -- 1990$^\dagger$ & 106 -- 956 & 3.4 -- 30.2 \\
        Subset Sample & 1.01 & 11.0 & 5.30 & 140.7 & 50.1 & 1857 $\pm$ 905 & $1528 \pm 793$ & $30 \pm 16$ \\
\enddata
    \caption{Median properties of the subsamples stacked in Sec. \ref{sec:subsamp}, and their respective upper limits on mass outflow rates and mass loading factor. Outflow velocities from the ionized gas population trends reported in \cite{fs19} are marked with a $\dagger$ symbol. In cases where outflow velocities are reported for both SF- and AGN-driven outflows, we use two velocity values corresponding to the average outflow velocity for these subsamples.}
    \label{tab:avg_prop}
\end{deluxetable*}

\subsection{The Gas Phase of Outflows}

The vast majority of cold molecular gas outflow detections are in the Local Universe, found in starbursts \citep{bolatto2013, fluetsch19}, (U)LIRGs \citep{lutz20, hc2020_2, hc2020_3, fluetsch2021} and AGNs (\citealt{cicone12, fiore2017, fluetsch19}; see \citealt{harrison2024} for a recent review). The reported outflow velocities range from a few tens of km/s \citep{fluetsch19} to $> 10^3$ km/s \citep{fiore2017, fluetsch19, lutz20}, although most have velocities of the order $10^2$ km/s \citep{walter2002, bolatto2013b, fiore2017, fluetsch19, lutz20, krieger2019}. In line with the nearby Universe, the few CO detections at cosmic noon have similar velocities: \cite{brusa2018} report an outflow detected with velocity $\sim 700$ km/s, \cite{hc19} find velocities between $\sim 300 - 500$ km/s, and the quasars in \cite{vayner2021} host outflows at $\sim 400 - 1100$ km/s. In general, all studies find that molecular gas outflow strength scales with AGN luminosity when one is present. The bulk of the mass budget in the outflows is dominated by the molecular gas phase, although there are no CO detections of SF-driven outflows at cosmic noon to compare with (see also Langan et al. in prep). In contrast, outflows in warmer gas phases, such as ionized gas, display systematically higher velocities: for AGN-driven outflows, velocities are $\sim 1000 - 2000$ km/s, against a few $\sim 100$ km/s for SF-driven outflows \citep{fluetsch19, fs19, lutz20}. With these differences in outflow properties depending on gas phase and power source mechanism, it is not surprising that detecting molecular gas outflows in typical SFGs and low luminosity AGNs at cosmic noon remains a challenge.


However, the challenging nature of the detection may not be the only explanation for the lack of outflow detections in this study. Indeed, the cold molecular phase of outflows might not be the dominant one: most of the outflow could be in a warmer molecular phase as a consequence of energy injection by stellar/AGN feedback in the galactic ISM. In the nearby Universe, studies have already investigated the hot molecular gas phase of outflows ($\gtrsim 1000$K; e.g., \citealt{ric2014,ra2019,riffel2023}) through observations of H$_2$ rovibrational lines, finding that the AGN could drive tens of Solar masses per year of warm molecular gas, which is unlikely to affect large-scale properties of the host galaxy \citep{ric2014}. More recently, \textit{JWST}/MIRI has opened the door to the study of warm molecular gas at $\sim 100 - 1000$K through purely rotational H$_2$ lines in the mid-IR, finding low ($\sim 150$ km/s) outflow velocities \citep{ric24, yuidan2024}. In some cases, the molecular gas outflows probed in these local AGNs come from the intersection of the ionised gas outflow with the galactic disk, which perturbs the gas within the galaxy but will not expel it from the galaxy \citep{ra2022, ric24}.

The H$_2$ molecules in the gas could also be dissociated by the galaxies' radiation fields into atomic hydrogen, particularly in MS galaxies. Using a simple theoretical model, \cite{vijayan2024} showed that the dissociation of molecular to atomic gas was mostly the result of radiative processes. Consequently, molecular gas in outflows can survive in starburst galaxies, thanks to the combination of denser environments and shorter dynamical times, which allows the molecular gas to escape the galaxy's radiation field before getting dissociated. In more typical star-forming galaxies, however, the molecules get dissociated, and the cool phase of outflows is mostly composed of atomic gas. In this scenario, the absence of cold molecular gas outflow signatures in our sample might be explained by the fact that the majority of the outflow is in the atomic phase.

The neutral atomic phase of outflows has also been observed, using low-ionization interstellar absorption lines in the rest-UV and optical (e.g., Mg II; Na I D), both in the local Universe (e.g., \citealt{rubin2014, zhu2015, cazzoli2016, rupke2018, saintonge2019, avery2022}), and at higher redshift  (e.g., \citealt{shapley2003, steidel2010, erb2012, sugahara2017}). Specifically, using facilities such as \textit{JWST} or Keck/LRIS, studies find prevalent neutral gas outflow signatures in cosmic noon galaxies \citep{fluetsch19, fluetsch2021, kehoe2024, rebecca24, belli2024, taylor2024}. In particular, \cite{belli2024} find a significant neutral outflow component with velocity $\sim 200$ km/s in a post-starburst galaxy at $z$ = 2.45, which could explain the rapid quenching of star formation in the galaxy. In line with this finding, \cite{rebecca24} observe neutral outflows through Na I D absorption in $z$ = $1.7 - 3.5$ galaxies, with velocities $200 - 1100$ km/s, which are found in star-forming and quenching systems alike. In a similar ($1.7 < z < 2.7$) redshift range, \cite{kehoe2024} report neutral outflows detected through low ionization interstellar absorption lines, albeit with lower mean velocities ($\langle \Delta v_{LIS} \rangle = -56$ km/s). These results indicate that, while the exact properties and impact on galaxy evolution are still uncertain, the neutral gas phase of outflows might play a significant role in the evolution of their host galaxy and represent a major fraction of the outflowing material.

Going along in the direction of these recent observations and theoretical models, \cite{rathjen2023} investigate the gas phase of stellar feedback using state-of-the-art magneto-hydrodynamic simulations, where non-equilibrium chemical network allows one to look into gas at temperatures down to $\sim 5$ K. They find that $50 - 90 \%$ of the outflowing material is in the so-called warm phase, defined as gas between $300 - 10^4$ K, which encompasses gas from the warm molecular gas phase to the ionised gas. As observations find that ionised gas does not dominate the mass budget of outflows \citep{freeman19, fs19, rebecca2020}, and in light of the recent studies discussed previously, this result seems in line with the picture in which gas phases other than the cold molecular and hot ionised gas are major contributors of the total mass budget of outflows.

\section{Conclusion}\label{sec:conc}
We present the results from a stacking analysis of the PHIBSS sample, for 154 galaxies at $0.5 \leq z \leq 2.6$, aiming to detect broad wings that would signal the presence of fast-moving outflowing material. We analyse the full sample and physically motivated sub-samples: above and below log$(M_\star) = 10.7 \: M_\odot$, above and below $\Delta$MS $= 0.2$, above and below $z = 1.7$, with inclination below 30\textdegree, between 30\textdegree and 60\textdegree, and above 60\textdegree, with $\mathrm{sSFR} > 0.1 \: \mathrm{Gyr}^{-1}$, and identified as AGN. We also perform a bootstrap-like subset sampling analysis to search for a sub-sample with a maximized broad component detection. Reaching integrated SNR  $> 30$, \textbf{we observe no outflow signatures in the full sample or in any of the physically motivated sub-samples (Sec. \ref{sec:fit}) and a 4$\sigma$ detection in the subset sampling analysis (Sec. \ref{sec:subsamp}) that could correspond to an outflow signature}. This latter statistically-identified subset represents 26\% of the full sample and does not show a clear differentiation in galaxy properties compared to the full sample.

Several factors could impact our ability to recover cold molecular gas outflows despite the use of spectral stacking: 1) a low outflow amplitude and/or low velocities, 2) a low outflow incidence, such that stacking in a larger sample could not reveal their presence (in line with the tentative detection when stacking only a quarter of the full sample), 3) a precise localization of the outflows in the galaxy, such that unresolved studies cannot detect them. The latter two are related to dilution effects of the broad component in a large sample and/or regions where emission is strongly dominated by gravitationally bound gas within the galaxies. However, we note that surface star formation density $\Sigma_\mathrm{SFR}$ of the PHIBSS sample follows a similar distribution to that of KMOS3D, where they detect ionized gas outflows through the stacking of the integrated galaxy spectra \citep{fs19}. Because of that, we do not expect the absence of molecular gas outflows to be the result of very diluted feedback due to low $\Sigma_\mathrm{SFR}$. On the other hand, the spatial resolution (and size of the extraction apertures) of our data probes physical scales of $\sim20$ kpc, and therefore, our analysis is limited to integrated galaxy-wide properties. Since molecular gas outflows are often more localized phenomena \citep{genzel2014, fs19}, which can extend to scales down to a few kpc or less \citep{veilleux17, hc19}, it is possible that we do not resolve molecular outflows present on smaller physical scales.

Derived upper limits on the mass outflow rate $\dot{M}_{out}$ and mass loading factor $\eta$ show that molecular outflows beneath our noise limits could still be more efficient than SF in depleting the gas. We stress that the computed upper limits are very conservative and rely on assumptions (such as the $\alpha_{CO}$) that may affect the value by factors of $> 5$.

In addition, in light of recent searches in alternative tracers at $z \sim 1 - 3$, the dominant phase of outflows in MS galaxies at cosmic noon might not be the cold molecular gas but rather warm molecular gas or neutral atomic gas. However, while outflows in those phases have been detected, there is not currently enough evidence to support either of those phases as the main outflow contributor. To evaluate the mass budget of outflows between the different gas phases at cosmic noon, deeper and higher resolution CO observations, along with observations of various gas tracers, will be needed. 

\begin{acknowledgments}
We thank the referee for their valuable and pertinent feedback, which helped us improve the paper. C.B., N.M.F.S., G.T., J.C., J.M.E.S. acknowledge funding by the European Union (ERC, GALPHYS, 101055023). HÜ acknowledges support through the ERC Starting Grant 101164796 ``APEX". Views and opinions expressed are, however, those of the author(s) only and do not necessarily reflect those of the European Union or the European Research Council. Neither the European Union nor the granting authority can be held responsible for them. This work is based on observations carried out with the IRAM PdBI/NOEMA interferometer. IRAM is supported by INSU/CNRS (France), MPG (Germany) and IGN (Spain).
We thank the referee for their valuable and pertinent feedback, which helped us improve the paper. C.B., N.M.F.S., G.T., J.C., J.M.E.S. acknowledge funding by the European Union (ERC, GALPHYS, 101055023). HÜ acknowledges support through the ERC Starting Grant 101164796 ``APEX". Views and opinions expressed are, however, those of the author(s) only and do not necessarily reflect those of the European Union or the European Research Council. Neither the European Union nor the granting authority can be held responsible for them. This work is based on observations carried out with the IRAM PdBI/NOEMA interferometer. IRAM is supported by INSU/CNRS (France), MPG (Germany) and IGN (Spain).
\end{acknowledgments}
%

\facilities{PdBI, NOEMA}


\software{Astropy \citep{2013A&A...558A..33A,2018AJ....156..123A}, 
          {\scshape SExtractor} \citep{1996A&AS..117..393B},
          {\scshape LineStacker} \citep{jolly2020}
          }



\appendix

\section{Sample Table} \label{ap:sample}

\begin{figure*}
    \centering
    \includegraphics[width=\linewidth]{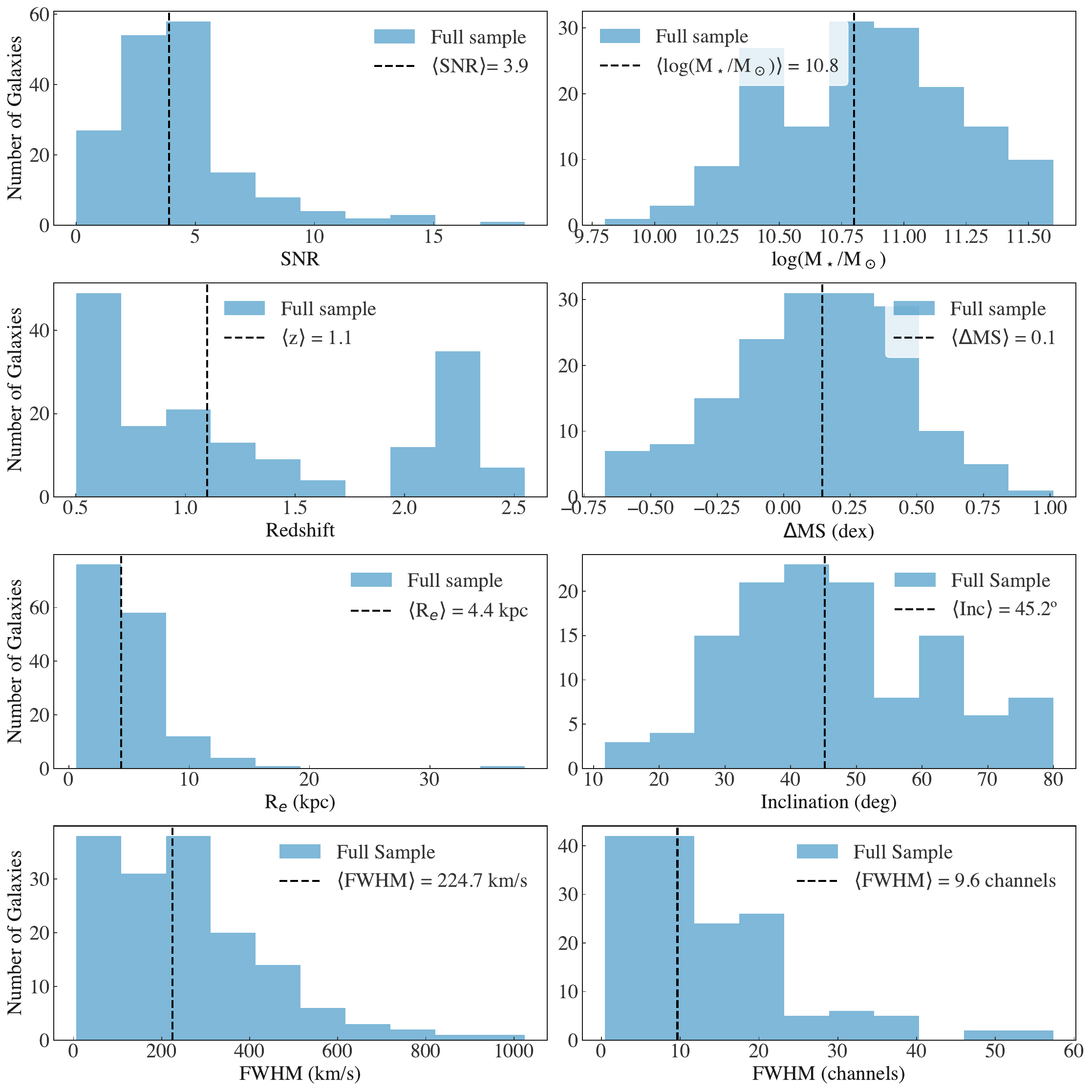}
    \caption{Distributions of the full sample's properties. The black dashed line in each panel indicates the median value. The FWHM distribution is plotted twice, in channels and km/s. \label{fig:sample_prop}}
\end{figure*}

This section presents the sample properties for the PHIBSS galaxies used in this analysis with CO line emission detected at integrated SNR $>$ 1.5 (see Table \ref{tab:sample}). The table does not report the full PHIBSS sample, but only the galaxies with a spectroscopic redshift estimate and lines with SNR $> 1.5$. The reported channel size is the channel size after the re-binning process described in Sec. \ref{sec:re-binning}. In addition, we also present in Fig. \ref{fig:sample_prop} the distribution of sample properties, including stellar mass, $\Delta$MS, SNR, inclination, redshift and effective radius. We also plot the distribution of line FWHM, both in km/s and in channels.

\begin{longrotatetable}
\begin{deluxetable*}{lcclccccccccc}
\tablewidth{700pt}
\tablecaption{PHIBSS Sample used in the analysis presented here, including galaxy properties derived from ancillary data. The target names with a dagger are part of the sample stacked and investigated in Sec. \ref{sec:subsamp} \& Appendix \ref{ap:sample}}
\tabletypesize{\scriptsize}
\tablehead{
    \colhead{Target} & 
    \colhead{R.A.} & \colhead{Decl.} & 
    \colhead{Field} & \colhead{Line} &
    \colhead{z$_{CO}$} &
    \colhead{$\mathrm{log(M_\star/M_\odot)}$} & \colhead{$\Delta$MS} &
    \colhead{Type} & \colhead{$R_e$} & 
    \colhead{FWHM} & \colhead{SNR} &
    \colhead{Channel Size} \\ 
    \colhead{} & 
    \colhead{} & \colhead{} &
    \colhead{} & \colhead{} & 
    \colhead{} &
    \colhead{} & \colhead{dex} &
    \colhead{} & \colhead{kpc} & 
    \colhead{km/s} & \colhead{} &
    \colhead{km/s}
}
\startdata
    EGS13034339               & 14:20:21.900 & +53:02:22.2  & EGS     & CO(3-2) & 1.44    & 10.8      & 0.2      &      & 3.0  & 249  & 5.4   & 36      \\
    XL55                      & 12:37:10.56  & +62:11:40.7  & GOODS-N & CO(2-1) & 0.78802 & 10.5      & 0.25     &      & 6.6  & 385  & 1.2   & 55      \\
    EGS13018632$^\dagger$     & 14:19:49.100 & +52:56:39.2  & EGS     & CO(3-2) & 1.229   & 10.7      & 0.25     &      & 1.9  & 221  & 3.73  & 32      \\
    XF53$^\dagger$            & 09:58:33.86  & +02:19:50.9  & COSMOS  & CO(2-1) & 0.50198 & 11.1      & 0.08     &      & 6.9  & 546  & 2.08  & 78      \\
    L14GN008                  & 12:36:07.83  & +62:12:00.6  & GOODS-N & CO(2-1) & 0.50276 & 10.3      & -0.27    &      & 6.4  & 160  & 3.01  & 23      \\
    L14CO004                  & 10:00:40.29  & +02:20:32.6  & COSMOS  & CO(2-1) & 0.68822 & 10.5      & -0.15    &      &      & 360  & 3.5   & 51      \\
    EGS13034541               & 14:20:22.200 & +53:02:14.8  & EGS     & CO(3-2) & 1.442   & 10.97     & 0.51     &      & 8.0  & 435  & 6.33  & 62      \\
    EGS13011148               & 14:19:29.810 & 52:53:13.130 & EGS     & CO(3-2) & 1.173   & 11.0      & -0.1     &      & 5.5  & 384  & 4.69  & 55      \\
    L14GN014                  & 12:36:51.82  & +62:15:04.7  & GOODS-N & CO(3-2) & 2.19044 & 10.6      & 0.23     & AGN  & 2.0  & 283  & 1.23  & 40      \\
    L14EG013$^\dagger$        & 14:19:17.33  & +52:50:35.3  & EGS     & CO(2-1) & 0.65855 & 11.1      & 0.48     & AGN  & 8.5  & 353  & 4.81  & 50      \\
    XA55                      & 12:36:59.92  & +62:14:50.0  & GOODS-N & CO(2-1) & 0.7609  & 10.5      & 0.65     &      & 4.0  & 126  & 3.35  & 18      \\
    EGS12028325               & 14:18:59.200 & +52:47:16.9  & EGS     & CO(3-2) & 1.159   & 10.4      & 0.43     &      & 3.3  & 178  & 5.81  & 25      \\
    L14GN035$^\dagger$        & 12:36:35.60  & +62:14:24.0  & GOODS-N & CO(6-5) & 2.01399 & 11.5      & 0.52     & AGN  & 7.7  & 616  & 2.54  & 88      \\
    XU53                      & 10:00:40.37  & +02:23:23.6  & COSMOS  & CO(2-1) & 0.51638 & 10.3      & 0.43     &      & 6.8  & 262  & 1.7   & 37      \\
    L14EG010                  & 14:20:22.80  & +52:55:56.3  & EGS     & CO(2-1) & 0.66944 & 10.7      & -0.13    &      & 2.6  & 27   & 10.11 & 4       \\
    L14EG011                  & 14:20:26.20  & +52:57:04.9  & EGS     & CO(2-1) & 0.57094 & 10.7      & 0.27     &      & 9.7  & 405  & 3.19  & 58      \\
    L14CO027                  & 10:00:45.00  & +02:07:05.1  & COSMOS  & CO(3-2) & 2.16887 & 10.3      & 0.45     &      & 3.6  & 100  & 5.4   & 14      \\
    L14CO026$^\dagger$        & 09:59:55.85  & +02:06:50.2  & COSMOS  & CO(3-2) & 2.18079 & 10.4      & 0.77     &      & 3.6  & 6    & 8.67  & 1       \\
    Q2343-MD59                & 23:46:26.900 & 12:47:39.870 & other   & CO(3-2) & 2.011   & 10.9      & -0.66    &      & 7.4  & 216  & 3.02  & 31      \\
    L14EG018                  & 14:19:38.67  & +52:51:38.8  & EGS     & CO(6-5) & 2.325   & 11.5      & 0.42     &      & 4.5  & 56   & 2.89  & 8       \\
    XI55                      & 12:36:17.33  & +62:12:12.8  & GOODS-N & CO(3-2) & 1.01779 & 11.2      & -0.48    &      & 5.8  & 102  & 2.09  & 15      \\
    XI54$^\dagger$            & 14:19:40.94  & +52:51:57.2  & EGS     & CO(3-2) & 1.0125  & 11.5      & -0.67    &      & 7.8  & 337  & 2.59  & 48      \\
    HDF-BX1439                & 12:36:53.660 & 62:17:24.270 & GOODS-N & CO(3-2) & 2.187   & 10.8      & -0.1     &      & 4.1  & 279  & 1.81  & 40      \\
    EGS13011155$^\dagger$     & 14:19:41.600 & +52:52:56.5  & EGS     & CO(3-2) & 1.012   & 11.1      & 0.46     & AGN  & 7.8  & 470  & 5.37  & 67      \\
    L14GN009                  & 12:36:18.50  & +62:09:03.5  & GOODS-N & CO(3-2) & 1.67551 & 11.4      & -0.6     &      & 4.2  & 97   & 6.5   & 14      \\
    L14GN021$^\dagger$        & 12:36:03.26  & +62:11:11.0  & GOODS-N & CO(2-1) & 0.63836 & 10.7      & 0.77     &      & 1.2  & 373  & 7.8   & 53      \\
    L14GN020                  & 12:36:18.29  & +62:08:19.6  & GOODS-N & CO(2-1) & 1.01748 &           &          &      &      & 277  & 2.92  & 40      \\
    L14CO005$^\dagger$        & 10:00:28.70  & +02:17:45.4  & COSMOS  & CO(3-2) & 2.09946 & 11.3      & -0.38    & AGN  & 2.1  & 181  & 2.6   & 26      \\
    XD54                      & 14:19:46.35  & +52:54:37.2  & EGS     & CO(2-1) & 0.75471 & 10.4      & 0.49     &      & 3.2  & 207  & 5.03  & 30      \\
    XD55                      & 12:36:21.04  & +62:12:08.5  & GOODS-N & CO(2-1) & 0.77927 & 10.5      & 0.25     &      &      & 218  & 3.1   & 31      \\
    EGS12012083               & 14:17:56.790 & 52:32:00.290 & EGS     & CO(3-2) & 1.119   & 11.0      & 0.3      &      & 4.4  & 576  & 3.67  & 82      \\
    EGS12023832               & 14:19:06.100 & +52:43:12.3  & EGS     & CO(3-2) & 1.351   & 10.8      & 0.4      &      & 4.7  & 395  & 2.26  & 56      \\
    L14CO012                  & 10:00:45.53  & +02:33:39.6  & COSMOS  & CO(2-1) & 0.70074 & 10.6      & -0.09    &      & 4.4  & 53   & 5.61  & 8       \\
    EGS13019114$^\dagger$     & 14:19:41.100 & +52:56:16.3  & EGS     & CO(3-2) & 1.105   & 10.8      & -0.0     &      & 7.2  & 419  & 6.38  & 60      \\
    L14CO016$^\dagger$        & 10:00:11.16  & +02:35:41.6  & COSMOS  & CO(2-1) & 0.69648 & 11.0      & 0.2      &      & 3.4  & 228  & 2.71  & 33      \\
    XK53                      & 10:01:59.05  & +01:46:58.1  & COSMOS  & CO(3-2) & 2.024   & 10.5      & 0.2      &      & 1.4  & 29   & 8.83  & 4       \\
    L14GN003                  & 12:36:11.52  & +62:10:33.6  & GOODS-N & CO(3-2) & 2.24344 & 11.3      & 0.12     &      & 5.6  & 807  & 1.82  & 115     \\
    L14GN002                  & 12:36:44.83  & +62:17:16.0  & GOODS-N & CO(3-2) & 2.03182 & 10.8      & 0.1      &      & 3.8  & 128  & 4.36  & 18      \\
    EGS12004280               & 14:17:00.900 & +52:27:01.3  & EGS     & CO(3-2) & 1.023   & 10.6      & 0.41     &      & 4.7  & 211  & 3.5   & 30      \\
    Q1700-BX691               & 17:01:06.000 & 64:12:10.270 & other   & CO(3-2) & 2.189   & 10.9      & -0.36    &      & 4.0  & 208  & 2.25  & 30      \\
    XC53$^\dagger$            & 10:00:58.20  & +01:45:59.0  & COSMOS  & CO(2-1) & 0.61686 & 10.9      & 0.52     & AGN  &      & 118  & 2.78  & 17      \\
    Q1700-MD94                & 17:00:42.020 & 64:11:24.220 & other   & CO(3-2) & 2.333   & 11.2      & 0.17     &      & 2.8  & 468  & 6.86  & 67      \\
    L14EG007$^\dagger$        & 14:19:19.01  & +52:48:30.5  & EGS     & CO(2-1) & 1.5274  &           &          & AGN  &      & 71   & 3.41  & 10      \\
    L14EG006                  & 14:18:45.52  & +52:43:24.1  & EGS     & CO(2-1) & 0.50148 & 10.5      & -0.15    &      & 8.0  & 71   & 8.34  & 10      \\
    L14CO018                  & 10:00:58.20  & +01:45:59.0  & COSMOS  & CO(2-1) & 0.61686 & 10.9      & 0.52     & AGN  &      & 515  & 3.07  & 74      \\
    L14GN015                  & 12:36:43.19  & +62:11:48.0  & GOODS-N & CO(2-1) & 1.00996 & 10.9      & -0.45    &      & 5.3  & 216  & 3.66  & 31      \\
    Q1700-MD174               & 17:00:54.540 & 64:16:24.760 & other   & CO(3-2) & 2.34    & 11.4      & -0.13    &      & 4.0  & 834  & 3.3   & 119     \\
    L14CO003                  & 10:00:43.81  & +02:14:09.2  & COSMOS  & CO(3-2) & 2.18195 & 10.6      & 0.33     &      & 3.1  & 36   & 6.23  & 5       \\
    L14CO002                  & 10:00:16.43  & +02:23:00.8  & COSMOS  & CO(3-2) & 2.18422 & 10.5      & 0.6      &      & 1.3  & 89   & 3.49  & 13      \\
    XD53$^\dagger$            & 10:01:58.73  & +02:15:34.2  & COSMOS  & CO(2-1) & 0.70203 & 10.9      & 0.42     &      & 3.4  & 393  & 5.17  & 56      \\
    EGS12020405               & 14:18:04.990 & 52:40:25.290 & EGS     & CO(3-2) & 1.379   & 10.6      & 0.61     &      & 7.4  & 437  & 2.79  & 62      \\
    Q1623-BX528               & 16:25:56.440 & 26:50:15.440 & other   & CO(3-2) & 2.268   & 10.8      & -0.2     &      & 4.6  & 56   & 3.64  & 8       \\
    L14GN030                  & 12:36:25.30  & +62:10:35.6  & GOODS-N & CO(3-2) & 2.08236 & 11.0      & -0.12    &      & 3.8  & 499  & 5.53  & 71      \\
    L14GN018                  & 12:36:31.66  & +62:16:04.1  & GOODS-N & CO(2-1) & 0.78317 & 10.4      & 0.49     &      & 3.4  & 179  & 5.47  & 26      \\
    HDF-BX1439$^\dagger$      & 14:19:09.500 & 52:53:06.400 & EGS     & CO(3-2) & 1.099   & 10.9      & 0.05     &      & 2.4  & 111  & 5.69  & 16      \\
    Q1700-BX561               & 17:01:04.180 & 64:10:43.830 & other   & CO(3-2) & 2.434   &           &          &      & 1.0  & 235  & 2.25  & 34      \\
    L14GN004                  & 12:37:04.34  & +62:14:46.2  & GOODS-N & CO(3-2) & 2.2143  & 10.7      & 0.37     & AGN  & 0.7  & 296  & 4.22  & 42      \\
    L14GN005$^\dagger$        & 12:37:20.05  & +62:12:22.8  & GOODS-N & CO(3-2) & 2.46026 & 10.8      & 0.4      &      & 2.8  & 144  & 4.24  & 21      \\
    L14CO021$^\dagger$        & 10:00:24.70  & +02:29:12.1  & COSMOS  & CO(2-1) & 0.70155 & 11.5      & 0.07     &      & 2.5  & 334  & 2.71  & 48      \\
    XK55                      & 12:36:46.19  & +62:11:42.1  & GOODS-N & CO(3-2) & 1.01578 & 11.4      & -0.25    &      & 9.8  & 284  & 2.7   & 41      \\
    XC55                      & 12:36:09.76  & +62:14:22.6  & GOODS-N & CO(2-1) & 0.78013 & 10.7      & 0.37     &      & 2.9  & 289  & 5.91  & 41      \\
    XC54$^\dagger$            & 14:19:49.14  & +52:52:35.8  & EGS     & CO(2-1) & 0.50933 & 11.3      & 0.37     &      & 14.6 & 688  & 2.12  & 98      \\
    EGS13011166$^\dagger$     & 14:19:45     & +52:52:28.0  & EGS     & CO(3-2) & 1.529   & 11.1      & 0.07     &      & 6.5  & 14   & 5.63  & 2       \\
    XW53                      & 10:00:45.52  & +02:16:34.3  & COSMOS  & CO(2-1) & 0.74949 & 10.4      & 0.09     &      &      & 89   & 2.53  & 13      \\
    L14GN019$^\dagger$        & 12:36:29.02  & +62:09:48.1  & GOODS-N & CO(3-2) & 2.31475 &           &          &      &      & 40   & 2.84  & 6       \\
    EGS13034445               & 14:20:30.800 & 53:01:48.500 & EGS     & CO(3-2) & 1.168   & 11.1      & -0.34    &      & 3.8  & 299  & 3.29  & 43      \\
    XL53$^\dagger$            & 10:00:28.27  & +02:16:00.5  & COSMOS  & CO(2-1) & 0.74848 & 11.2      & 0.27     &      & 1.6  & 42   & 6.17  & 6       \\
    XR53                      & 10:01:41.85  & +02:07:09.8  & COSMOS  & CO(2-1) & 0.51685 & 11.3      & -0.03    &      &      & 176  & 4.01  & 25      \\
    L14GN026                  & 12:36:36.74  & +62:17:47.8  & GOODS-N & CO(3-2) & 2.21233 &           &          &      &      & 26   & 9.11  & 4       \\
    XF55$^\dagger$            & 12:35:55.43  & +62:10:56.8  & GOODS-N & CO(2-1) & 0.6381  & 10.2      & 0.08     &      & 6.9  & 408  & 2.76  & 58      \\
    XF54$^\dagger$            & 14:19:41.70  & +52:55:41.3  & EGS     & CO(2-1) & 0.76832 & 10.8      & 0.14     &      & 6.2  & 218  & 3.64  & 31      \\
    XA53$^\dagger$            & 10:02:02.09  & +02:09:37.4  & COSMOS  & CO(2-1) & 0.69861 & 11.5      & 0.47     &      & 10.1 & 458  & 4.91  & 65      \\
    L14GN013                  & 12:37:00.46  & +62:15:08.9  & GOODS-N & CO(3-2) & 2.32941 & 11.1      & -0.18    & AGN  & 1.8  & 282  & 3.91  & 40      \\
    L14GN012$^\dagger$        & 12:37:07.20  & +62:14:08.1  & GOODS-N & CO(3-2) & 2.48614 & 11.3      & -0.08    & AGN  & 4.0  & 539  & 6.39  & 77      \\
    Q1623-BX599               & 16:26:02.540 & 26:45:31.900 & other   & CO(3-2) & 2.33    & 10.8      & 0.1      &      & 1.0  & 98   & 7.13  & 14      \\
    EGS13004661               & 14:18:40.840 & 52:48:35.650 & EGS     & CO(3-2) & 1.192   & 10.5      & 0.27     &      & 6.3  & 496  & 1.98  & 71      \\
    L14EG016                  & 14:18:28.90  & +52:43:05.3  & EGS     & CO(2-1) & 0.64418 &           &          &      &      & 228  & 2.77  & 33      \\
    EGS13004684$^\dagger$     & 14:18:55.800 & +52:47:49.5  & EGS     & CO(3-2) & 1.014   & 11.0      & -0.2     &      & 5.0  & 353  & 2.63  & 51      \\
    L14CO008$^\dagger$        & 09:58:09.07  & +02:05:29.8  & COSMOS  & CO(2-1) & 0.60674 & 10.9      & -0.08    &      &      & 196  & 5.39  & 28      \\
    L14CO020                  & 11:24:15.64  & -21:39:31.0  & COSMOS  & CO(3-2) & 2.38257 &           &          &      &      & 31   & 7.04  & 4       \\
    L14CO009                  & 09:58:56.45  & +02:08:06.7  & COSMOS  & CO(2-1) & 0.69763 & 10.5      & 0.25     &      & 5.8  & 273  & 3.65  & 39      \\
    EGS13004291               & 14:19:15     & +52:49:29.9  & EGS     & CO(3-2) & 1.145   & 10.97     & 1.01     & AGN  & 3.1  & 335  & 23.17 & 48      \\
    XE55                      & 12:36:11.26  & +62:14:20.9  & GOODS-N & CO(2-1) & 0.76796 & 10.5      & 0.15     &      & 5.2  & 176  & 4.4   & 25      \\
    XE54                      & 14:19:35.27  & +52:52:49.9  & EGS     & CO(2-1) & 0.50945 & 10.4      & -0.01    &      & 8.2  & 277  & 4.71  & 40      \\
    EGS13017614               & 14:20:24.300 & 52:55:41.300 & EGS     & CO(3-2) & 1.18    & 11.1      & 0.06     &      & 5.9  & 360  & 8.26  & 51      \\
    Q2346-BX482               & 23:48:12.969 & 00:25:46.336 & other   & CO(3-2) & 2.264   & 9.8       & 0.26     &      & 3.8  & 185  & 3.86  & 26      \\
    EGS12011767               & 14:18:24.710 & 52:32:55.250 & EGS     & CO(3-2) & 1.282   & 10.5      & 0.37     &      & 6.0  & 133  & 6.98  & 19      \\
    L14CO019$^\dagger$        & 10:00:35.69  & +02:31:15.6  & COSMOS  & CO(2-1) & 0.6777  & 10.9      & 0.22     &      & 14.9 & 122  & 3.54  & 17      \\
    EGS13042293               & 14:20:40.800 & +53:04:59.2  & EGS     & CO(3-2) & 1.393   & 10.6      & 0.11     &      & 5.2  & 195  & 6.11  & 28      \\
    EGS13003805$^\dagger$     & 14:19:40.070 & 52:49:38.560 & EGS     & CO(3-2) & 1.23    & 11.2      & 0.42     & AGN  & 5.6  & 409  & 9.12  & 58      \\
    XO53$^\dagger$            & 10:02:51.41  & +02:18:49.7  & COSMOS  & CO(2-1) & 0.60681 & 11.4      & -0.14    & AGN  & 1.1  & 194  & 3.41  & 28      \\
    EGS12004351               & 14:17:02     & +52:26:58.5  & EGS     & CO(3-2) & 1.017   & 10.9      & 0.45     &      & 5.7  & 537  & 6.77  & 77      \\
    L14GN016                  & 12:37:22.94  & +62:14:19.7  & GOODS-N & CO(2-1) & 1.02173 & 11.4      & -0.25    &      & 5.2  & 37   & 5.62  & 5       \\
    L14GN022                  & 12:36:36.76  & +62:11:56.1  & GOODS-N & CO(2-1) & 0.55611 & 10.1      & -0.06    & AGN  & 1.0  & 93   & 2.4   & 13      \\
    L14CO006                  & 10:00:31.08  & +02:12:25.9  & COSMOS  & CO(3-2) & 2.30902 & 10.6      & -0.17    &      & 4.1  & 477  & 3.43  & 68      \\
    L14CO022                  & 10:01:47.00  & +02:23:25.0  & COSMOS  & CO(3-2) & 2.208   & 10.3      & 0.45     &      & 2.0  & 80   & 5.26  & 11      \\
    EGS13019128               & 14:19:38.100 & +52:55:40.9  & EGS     & CO(3-2) & 1.35    & 10.6      & 0.31     &      & 5.2  & 194  & 7.77  & 28      \\
    Q2343-BX513               & 23:46:11.130 & 12:48:32.140 & other   & CO(3-2) & 2.109   & 10.4      & -0.33    &      & 4.0  & 134  & 2.2   & 19      \\
    XH55                      & 12:37:13.87  & +62:13:35.0  & GOODS-N & CO(2-1) & 0.77839 & 10.3      & 0.13     &      & 5.5  & 275  & 4.69  & 39      \\
    XH54                      & 14:19:45.42  & +52:55:51.0  & EGS     & CO(2-1) & 0.75573 & 10.2      & 0.18     &      & 5.4  & 74   & 3.91  & 11      \\
    X453                      & 16:25:50.850 & 26:49:31.300 & other   & CO(3-2) & 2.182   & 10.5      & 0.6      &      & 2.0  & 230  & 5.17  & 33      \\
    L14CO011                  & 10:00:14.30  & +02:30:47.2  & COSMOS  & CO(2-1) & 0.69685 & 10.4      & 0.49     &      & 1.9  & 286  & 7.53  & 41      \\
    L14EG005                  & 14:18:59.76  & +52:42:50.8  & EGS     & CO(3-2) & 2.16865 & 10.6      & 0.03     &      & 16.1 & 282  & 2.57  & 40      \\
    EGS13034542               & 14:20:21.200 & +02:05:04.2  & EGS     & CO(3-2) & 1.435   & 10.7      & 0.15     &      & 4.0  & 13   & 8.79  & 2       \\
    EGS13035123               & 14:20:05.500 & 53:01:15.600 & EGS     & CO(3-2) & 1.115   & 11.2      & 0.02     &      & 9.1  & 121  & 10.61 & 17      \\
    L14GN017$^\dagger$        & 12:36:53.66  & +62:17:24.3  & GOODS-N & CO(3-2) & 2.18711 & 10.5      & 0.2      &      & 3.4  & 180  & 2.96  & 26      \\
    EGS13017973               & 14:20:13.100 & +52:56:13.7  & EGS     & CO(3-2) & 1.031   & 10.6      & 0.11     &      & 7.2  & 78   & 7.62  & 11      \\
    L14GN029                  & 12:37:28.10  & +62:14:40.0  & GOODS-N & CO(3-2) & 2.54866 &           &          &      &      & 35   & 10.1  & 5       \\
    L14CO013                  & 10:02:16.78  & +01:37:25.0  & COSMOS  & CO(2-1) & 0.62109 & 11.2      & 0.07     & AGN  & 1.9  & 458  & 2.71  & 66      \\
    L14CO025                  & 09:59:57.20  & +02:12:25.2  & COSMOS  & CO(3-2) & 2.45796 & 11.0      & 0.08     &      &      & 86   & 6.77  & 12      \\
    L14EG012                  & 14:19:52.95  & +52:51:11.1  & EGS     & CO(2-1) & 0.54413 & 11.1      & -0.22    &      & 15.3 & 165  & 3.81  & 24      \\
    EGS13018076               & 14:20:10.800 & +52:53:41.5  & EGS     & CO(3-2) & 1.227   & 11.0      & 0.1      &      & 8.0  & 87   & 4.59  & 12      \\
    XT53                      & 10:01:39.31  & +02:17:25.8  & COSMOS  & CO(2-1) & 0.70096 & 11.1      & 0.18     &      &      & 265  & 3.72  & 38      \\
    L14CO010$^\dagger$        & 10:01:08.69  & +01:44:28.2  & COSMOS  & CO(3-2) & 2.23988 & 11.2      & 0.07     &      &      & 343  & 2.5   & 49      \\
    L14GN034$^\dagger$        & 12:36:19.68  & +62:19:08.1  & GOODS-N & CO(2-1) & 0.51939 & 10.9      & -0.28    &      & 9.6  & 753  & 2.94  & 108     \\
    EGS13018312               & 14:19:58.300 & +52:55:49.4  & EGS     & CO(3-2) & 1.105   & 10.8      & -0.2     &      & 4.0  & 251  & 6.99  & 36      \\
    L14GN023                  & 12:36:00.14  & +62:10:47.2  & GOODS-N & CO(3-2) & 1.99909 & 10.8      & 0.47     & AGN  & 0.6  & 237  & 2.06  & 34      \\
    L14CO007$^\dagger$        & 10:00:25.18  & +02:29:53.9  & COSMOS  & CO(2-1) & 0.50185 & 10.7      & -0.53    &      & 10.7 & 385  & 3.92  & 55      \\
    EGS12015684               & 14:18:42.090 & 52:36:20.160 & EGS     & CO(3-2) & 1.374   & 10.7      & 0.45     &      & 3.8  & 106  & 9.07  & 15      \\
    L14EG002                  & 14:19:27.42  & +52:47:55.6  & EGS     & CO(3-2) & 2.29574 & 11.0      & 0.18     &      & 1.9  & 99   & 3.77  & 14      \\
    EGS13026117               & 14:20:26.500 & +52:59:39.6  & EGS     & CO(3-2) & 1.241   & 11.1      & 0.26     &      & 3.2  & 168  & 20.87 & 24      \\
    L14EG003                  & 14:19:15.35  & +52:44:31.7  & EGS     & CO(3-2) & 2.02781 & 10.9      & 0.24     &      & 2.5  & 203  & 3.23  & 29      \\
    L14GN010$^\dagger$        & 12:36:41.42  & +62:11:42.5  & GOODS-N & CO(3-2) & 1.52663 & 11.2      & -0.49    &      & 6.8  & 699  & 6.32  & 100     \\
    L14GN011                  & 12:37:22.53  & +62:18:38.2  & GOODS-N & CO(3-2) & 1.52315 & 11.5      & -0.55    &      & 10.8 & 426  & 3.77  & 61      \\
    XM53                      & 10:01:53.57  & +01:54:14.8  & COSMOS  & CO(2-1) & 0.70033 & 11.6      & 0.17     &      & 2.1  & 648  & 3.28  & 93      \\
    EGS13033731               & 14:20:43.300 & +53:03:48.5  & EGS     & CO(3-2) & 1.317   & 10.4      & -0.07    &      & 5.5  & 87   & 2.42  & 12      \\
    L14GN006                  & 12:36:34.41  & +62:17:50.5  & GOODS-N & CO(2-1) & 0.68333 & 10.5      & 0.35     &      & 4.3  & 323  & 3.75  & 46      \\
    L14GN007                  & 12:36:32.38  & +62:07:34.1  & GOODS-N & CO(2-1) & 0.59423 & 10.8      & -0.26    &      & 6.1  & 441  & 4.59  & 63      \\
    EGS13017843               & 14:20:18.900 & +52:56:05.1  & EGS     & CO(3-2) & 1.052   & 10.6      & -0.09    &      & 4.2  & 518  & 3.35  & 74      \\
    L14CO023                  & 10:01:59.05  & +01:46:58.1  & COSMOS  & CO(3-2) & 2.024   & 10.5      & 0.2      &      & 1.4  & 57   & 6.06  & 8       \\
    XG54                      & 14:20:13.43  & +52:54:05.9  & EGS     & CO(2-1) & 0.65935 & 11.3      & -0.03    &      & 13.0 & 287  & 5.81  & 41      \\
    L14GN028                  & 12:37:02.93  & +62:14:23.6  & GOODS-N & CO(2-1) & 0.51138 & 10.8      & -0.26    &      & 5.0  & 278  & 5.37  & 40      \\
    L14EG014                  & 14:20:33.58  & +52:59:17.5  & EGS     & CO(2-1) & 0.71013 & 10.9      & -0.38    & AGN  & 4.5  & 119  & 3.59  & 17      \\
    L14EG015                  & 14:20:45.61  & +53:05:31.2  & EGS     & CO(2-1) & 0.73785 & 11.0      & -0.1     &      & 1.2  & 57   & 5.49  & 8       \\ 
    EGS12004754               & 14:16:42.100 & 52:25:19.100 & EGS     & CO(3-2) & 1.026   & 11.0      & -0.1     &      & 5.0  & 88   & 6.74  & 12      \\
    Q2343-BX610$^\dagger$     & 23:46:09.430 & 12:49:19.210 & other   & CO(3-2) & 2.211   & 11.0      & 0.18     &      & 4.0  & 256  & 15.26 & 37      \\
    Q2343-BX442               & 23:46:19.360 & 12:47:59.690 & other   & CO(3-2) & 2.175   & 11.1      & 0.02     &      & 2.0  & 291  & 2.3   & 42      \\
    L14GN033$^\dagger$        & 12:36:53.81  & +62:08:27.7  & GOODS-N & CO(2-1) & 0.56117 & 10.1      & -0.06    &      & 6.7  & 51   & 3.44  & 7       \\
    Q1700-MD69                & 17:00:47.620 & 64:09:44.780 & other   & CO(3-2) & 2.289   & 11.3      & -0.08    &      & 4.6  & 178  & 6.57  & 25      \\
    L14EG017                  & 14:19:49.28  & +52:51:34.1  & EGS     & CO(6-5) & 2.18691 & 10.9      & 0.24     & AGN  & 1.4  & 318  & 2.35  & 45      \\
    L14GN024                  & 12:37:23.47  & +62:17:20.2  & GOODS-N & CO(3-2) & 2.22332 & 10.6      & 0.23     &      & 4.2  & 114  & 4.23  & 16      \\
    L14CO001                  & 10:00:18.91  & +02:18:10.1  & COSMOS  & CO(2-1) & 0.50214 & 10.2      & 0.58     &      & 7.2  & 166  & 4.9   & 24      \\
    XE53                      & 10:01:00.74  & +01:49:53.0  & COSMOS  & CO(2-1) & 0.52892 & 10.4      & 0.39     &      & 2.0  & 240  & 5.16  & 34      \\
    EGS13017707$^\dagger$     & 14:20:13     & +52:55:34.0  & EGS     & CO(3-2) & 1.037   & 10.87     & 0.76     & AGN  & 3.6  & 304  & 5.39  & 43      \\
    EGS12024866               & 14:18:19.270 & 52:42:21.520 & EGS     & CO(3-2) & 1.002   & 10.4      & 0.03     &      & 4.4  & 215  & 3.71  & 31      \\
    L14GN025                  & 12:37:13.99  & +62:20:36.6  & GOODS-N & CO(2-1) & 0.53281 & 10.7      & -0.63    &      & 1.9  & 229  & 1.58  & 33      \\
    XJ55$^\dagger$            & 12:36:29.13  & +62:10:46.1  & GOODS-N & CO(3-2) & 1.01418 & 11.4      & 0.15     & AGN  & 8.5  & 1025 & 5.27  & 146     \\
    L14CO014                  & 10:01:09.67  & +02:30:00.7  & COSMOS  & CO(2-1) & 0.7019  & 10.7      & 0.17     &      & 0.6  & 119  & 4.66  & 17      \\
    XV53$^\dagger$            & 10:01:43.66  & +02:48:09.4  & COSMOS  & CO(2-1) & 0.62351 & 10.8      & 0.24     &      & 1.6  & 489  & 4.45  & 70      \\
    L14EG009$^\dagger$        & 14:20:04.88  & +52:59:38.8  & EGS     & CO(2-1) & 0.73586 & 10.1      & 0.14     &      & 3.0  & 211  & 3.66  & 30      \\
    L14EG008                  & 14:19:39.46  & +52:52:33.6  & EGS     & CO(2-1) & 0.7316  & 10.9      & 0.72     &      & 7.9  & 234  & 10.43 & 33      \\
    XB54                      & 14:19:37.26  & +52:51:03.4  & EGS     & CO(2-1) & 0.66997 & 11.4      & 0.26     & AGN  & 37.9 & 84   & 8.13  & 12      \\
    XB55                      & 12:36:08.13  & +62:10:35.9  & GOODS-N & CO(2-1) & 0.6791  &           &          & AGN  &      & 20   & 4.52  & 3      
\enddata
\label{tab:sample}
\end{deluxetable*}
\end{longrotatetable}

\section{Spectrum Extraction Process}\label{ap:spec_extr}
\begin{figure*}
    \centering
    \includegraphics[width=\linewidth]{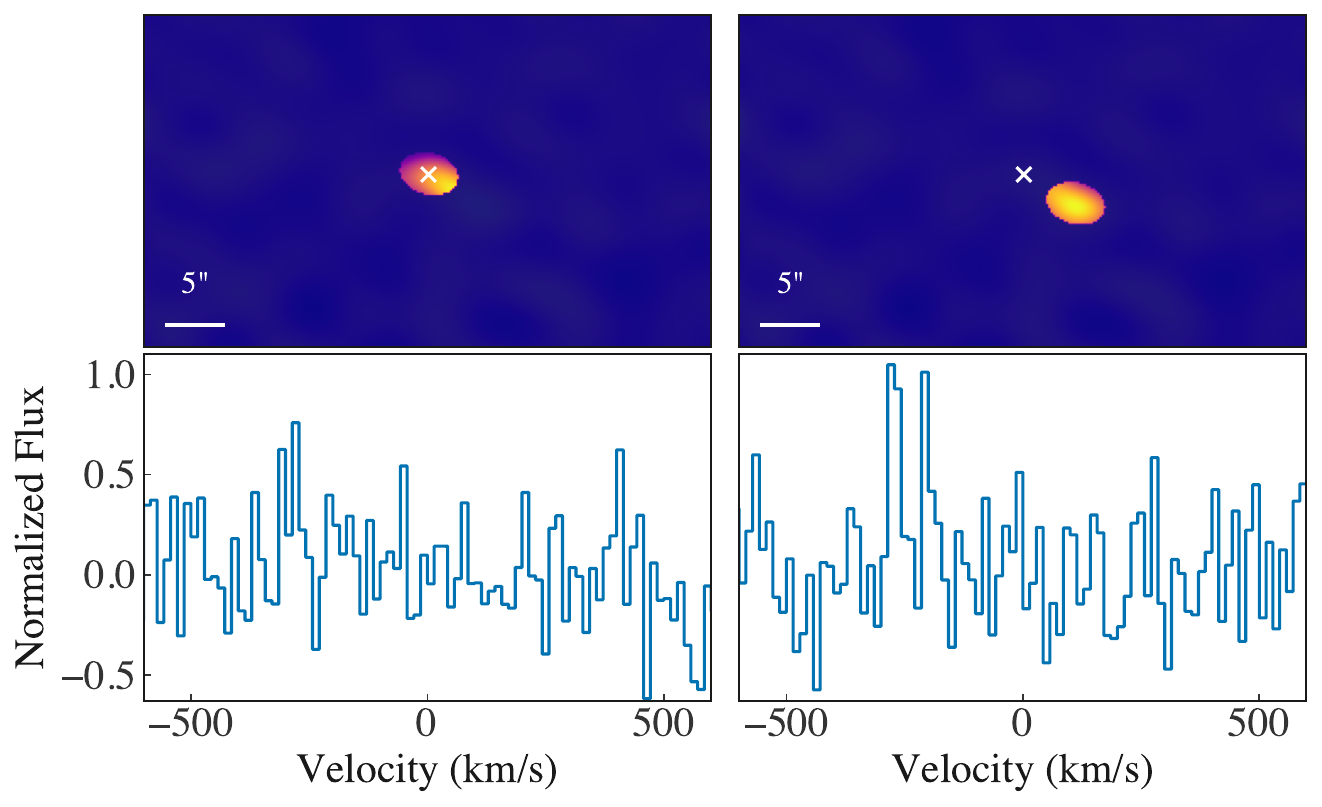}
    \caption{Example case in which the iterative process described in Sec. \ref{sec:spec} allowed to properly retrieve the spectrum of the galaxy, after correction the position of the source. \textit{Top row:} collapsed data cube, with the position of the beam-sized aperture used to extracted the spectrum from. The left panel shows the initial aperture used, whereas the right panel shows the position-corrected aperture, after running {\scshape SExtractor} and re-extracting the spectrum. In both panels, the white cross indicates the center of the frame. \textit{Bottom row:} Spectrum extracted from the aforementioned aperture. The right panel shows the detection of the line, which displays a double-peaked feature reminiscent of disk rotation in integrated spectra.}
    \label{fig:l14eg002}
\end{figure*}

As described in Sec. \ref{sec:spec}, all data cubes go through the same standardized iterative process to estimate as best as possible the position and spatial extent of the detection. Indeed, for some of the galaxies, the source is located away from the center of the field-of-view (indicated with a white cross on Fig. \ref{fig:l14eg002}). This effect is showcased in Fig. \ref{fig:l14eg002}, for L14EG002, a galaxy at $z = 2.3$, where the detection of the line is clearly affected by the position of the beam-sized aperture.

Similarly, collapsing the cube along an insufficient amount of spectral channels, or along the wrong ones, can lead to an inaccurate extraction of the spectra. The effect of this latter case is displayed in Fig. \ref{fig:5123}, Appendix \ref{ap:ind_spec}.

\section{Individual Spectra Investigation}\label{ap:ind_spec}
\begin{figure*}
    \centering
    \includegraphics[width=\linewidth]{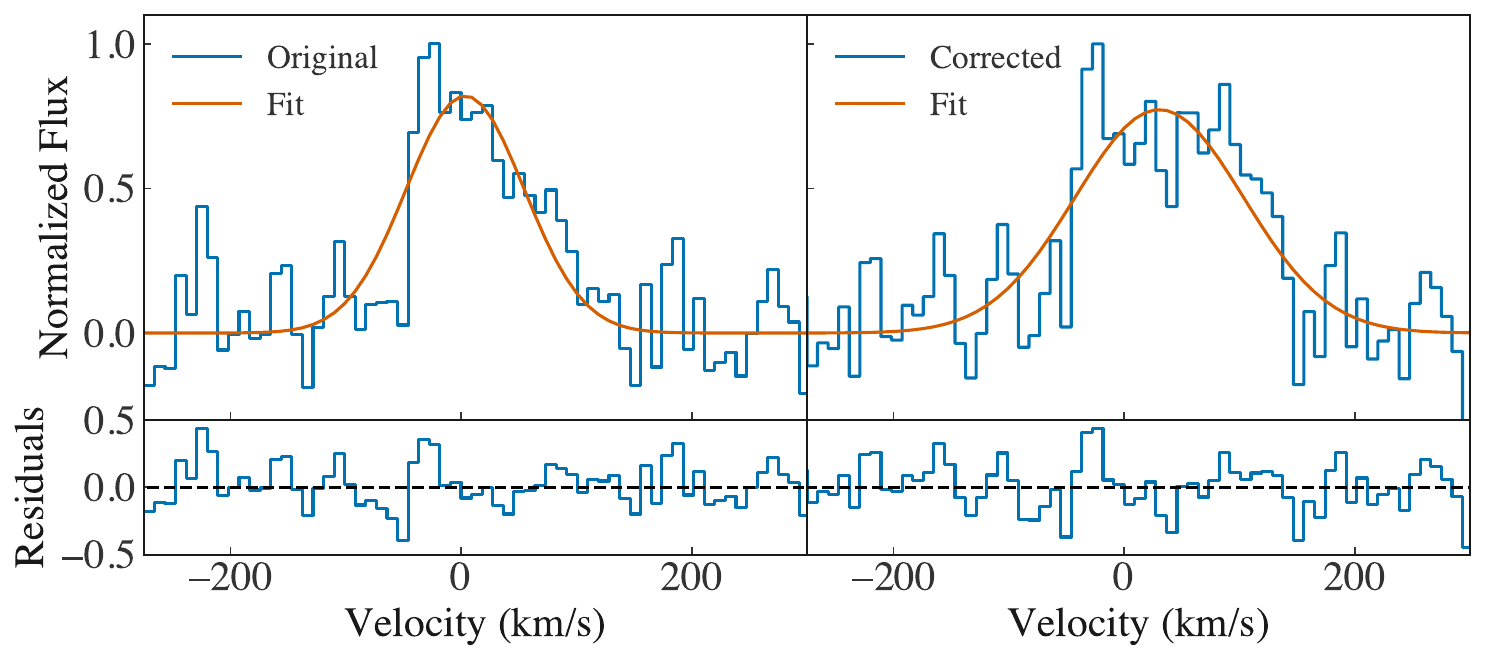}
    \caption{Example of a high SNR spectrum (blue curve) in EGS13035123 showing asymmetry in the CO(3-2) line profile (left panel), and how the line shape changes when correcting the aperture mask (right). Overlaid are the best-fit Gaussian profiles (orange lines), and the bottom panels show the fit residuals in both cases.}
    \label{fig:5123}
\end{figure*}

Three objects are discussed in Sec. \ref{sec:ind_spec}, as they display asymmetries in the line profile. For two of them (EGS13035123 and EGS12004351), correcting the mask used to extract the spectra is enough to change the asymmetry to the double-peaked profile indicative of galactic rotation. Fig. \ref{fig:5123} shows the spectra of one of these galaxies, EGS13035123, before and after adjusting the mask.

\begin{figure}
    \centering
    \includegraphics[width=\linewidth]{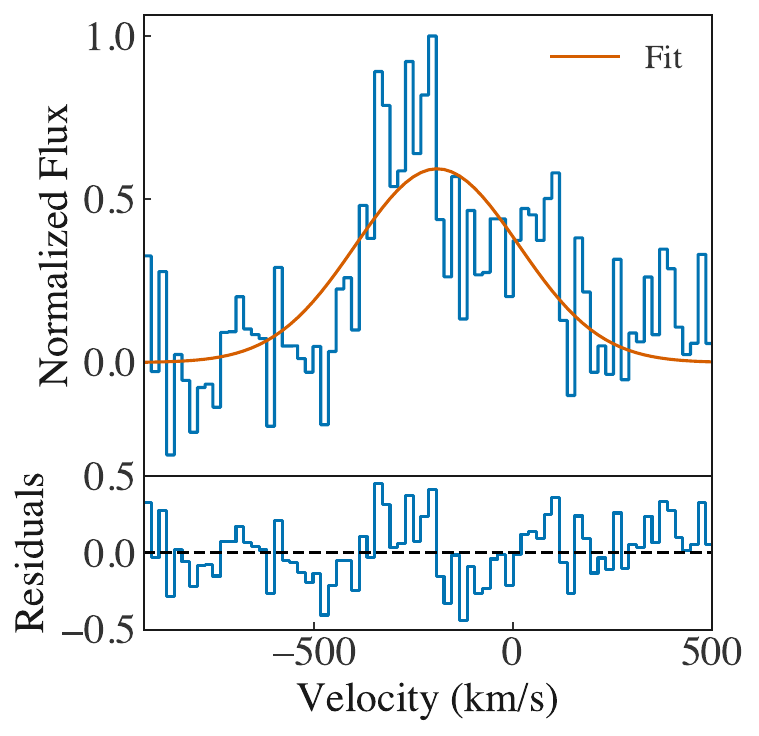}
    \caption{CO(3-2) spectrum of EGS13003805, the galaxy showing signs of asymmetry in the profile (blue curve), which cannot be corrected through the adjustment of the aperture. Overlaid is the best-fit Gaussian profile (orange line), and the bottom panel show the fit residuals.}
    \label{fig:3805}
\end{figure}

In contrast, in the case of EGS13003805 (Fig. \ref{fig:3805}), the spectrum retains its asymmetric shape even when changing the aperture. Based on the strength of the asymmetric feature and through visual inspection of the cube, we believe that the feature is still a signature of rotation. 

\section{Extended and Annuli Apertures}\label{ap:ap}
\begin{figure*}
    \centering
    \includegraphics[width=\linewidth]{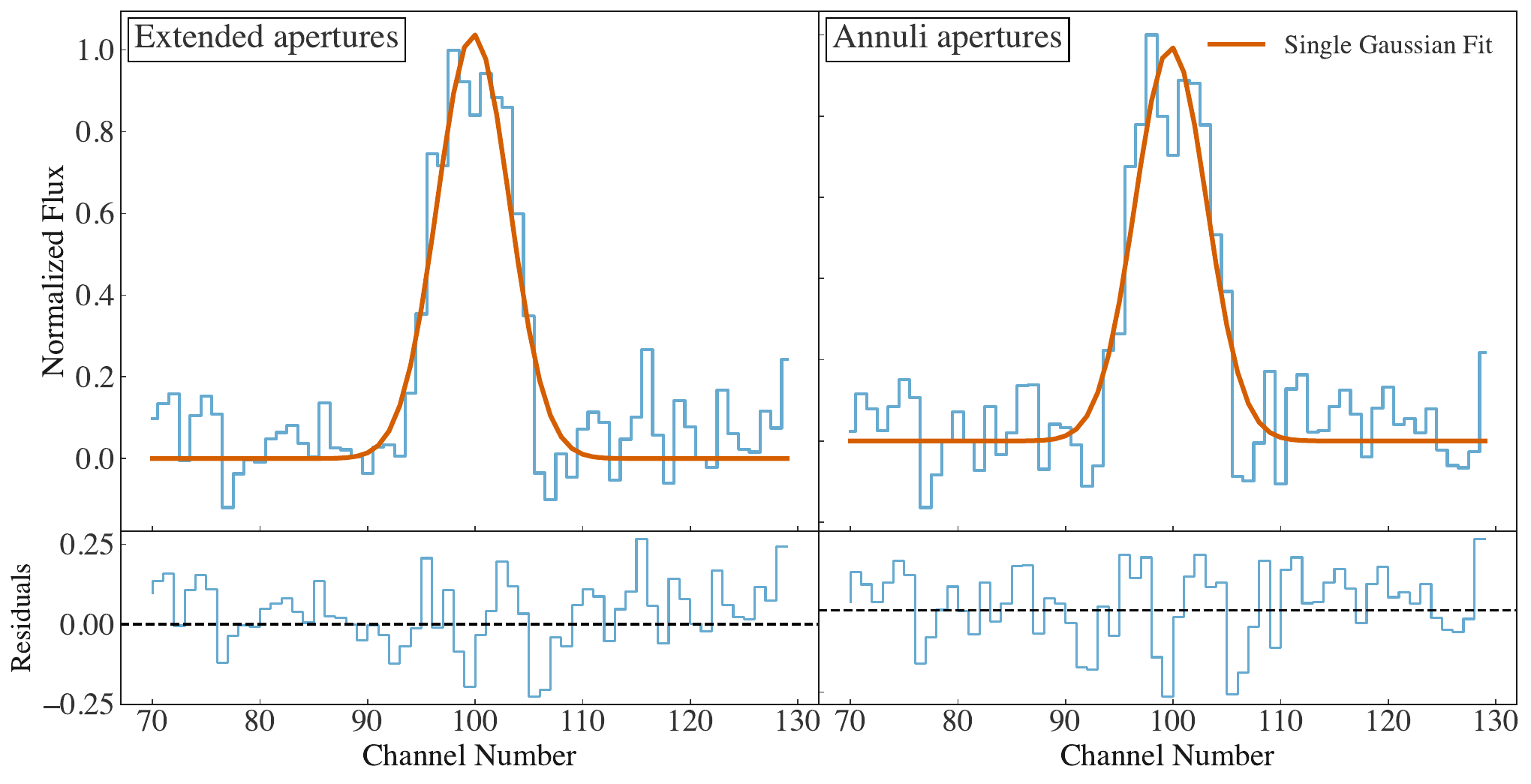}
    \caption{Stacked profiles for different apertures overlaid with the best-fit single Gaussian profile (solid red line). In both cases, fitting shows that no significant outflow component is detected in the stacked spectrum. \textit{Left Panel}: Top: Stacked spectrum (light blue curve) of the full sample of 154 galaxies, but extracted from apertures 10 kpc wider. Bottom: Residuals of the fit to the spectrum. {\textit{Right Panel}: Same as left, but extracted from 10 kpc wide annuli (the difference between the extended apertures and original apertures)} Bottom: Residuals of the fit to the spectrum.}
    \label{fig:big_ann_ap}
\end{figure*}

As described in Sec. \ref{sec:big_ap}, we performed the spectral stacking analysis not only on spectra extracted from the smallest possible aperture but also on spectra from apertures wider by 10 kpc (following the outflow detection in \citealt{hc19}), as well as on "annuli" apertures of only the additional 10kpc around the original aperture. The goal is to search extended emission in the galaxy and potential outflow signatures in the outskirts of the galaxies. The resulting spectra for the extended and annuli apertures are shown in Fig. \ref{fig:big_ann_ap}, along with the best-fit single Gaussian, and the fit residuals. Similar to Fig. \ref{fig:full_stack} and \ref{fig:stacks}, both the spectra and fit residuals show no indication of a broad secondary outflow component. Additional fitting of a double Gaussian model does not improve the residuals, or the goodness-of-fit (verified through a reduced $\chi^2$ and $\Delta$BIC analysis).

\section{Subset Sampling Investigation}\label{ap:subsamp}

\begin{figure*}[h!]
    \centering
    \includegraphics[width=\linewidth]{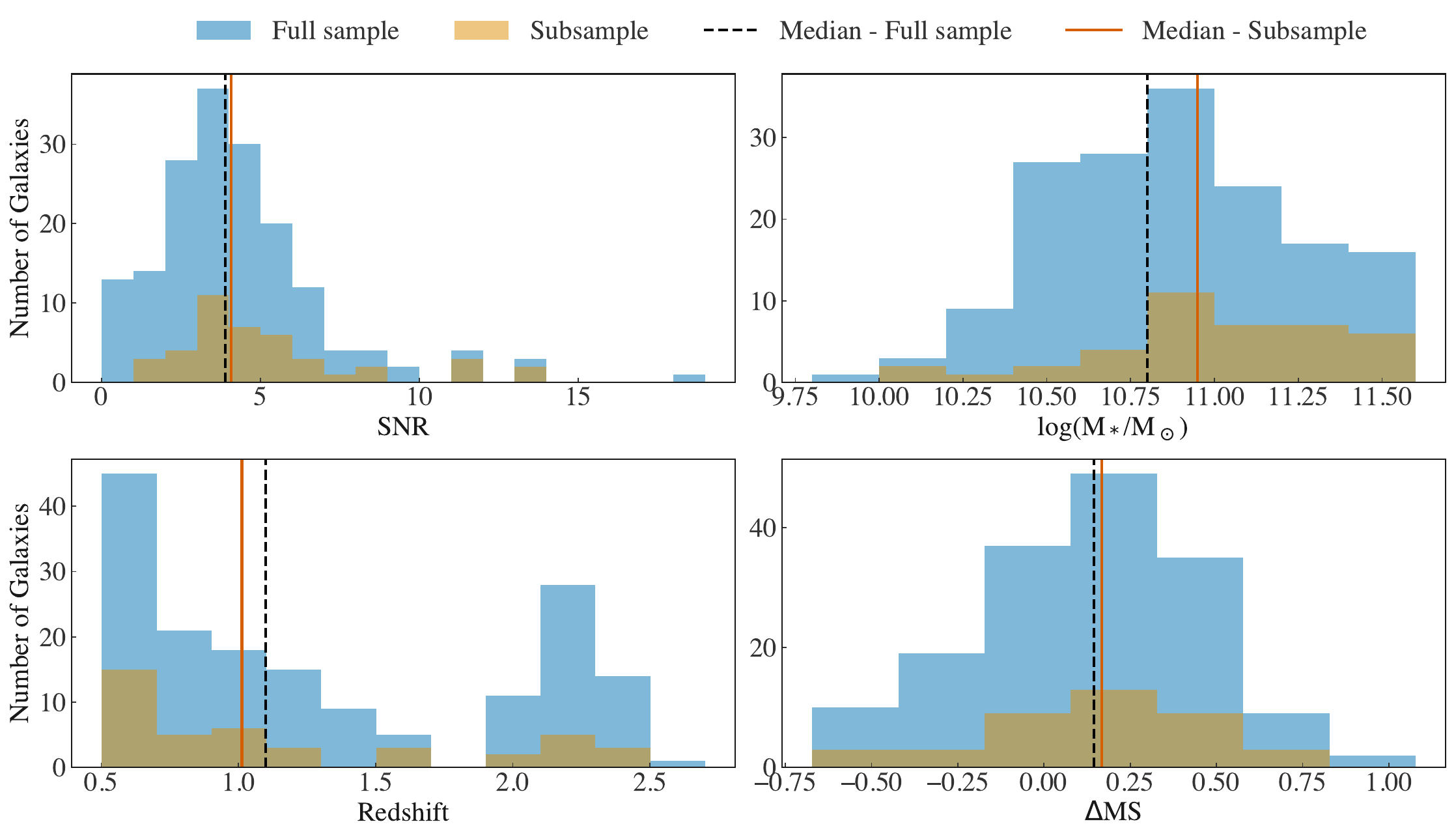}
    \caption{Comparison between properties distributions of the full sample (light blue histograms) and the subset sample (yellow histograms) showing a tentative outflow detection. The medians of each properties are also plotted for the full sample (dashed black) and the subset sample (solid orange).}
    \label{fig:subsamp_prop}
\end{figure*}

As part of the subset sampling analysis presented in Sec. \ref{sec:subsamp}, we investigate the subsample of 41 galaxies with the highest grade, which also presents signs of a tentative outflow detection (marked with a $\dagger$ symbol in Table \ref{tab:sample}). As such, we compare the galaxy properties of the subsample to that of the full sample (see Fig. \ref{fig:subsamp_prop}) and see that there are no clear distinctions in the distribution of data and galaxy properties. The SNR and $\Delta$MS distributions have almost the same median, and the subsample median is at a slightly lower redshift compared to that of the full sample (1.012 vs 1.099, respectively). The only noticeable difference in sample properties is the median stellar mass, which is $10^{10.8}$ M$_\odot$ for the full sample, and $10^{10.95}$ M$_\odot$ for the subsample. In addition, 13 of these 41 galaxies are identified as AGN (see Sec. \ref{sec:agn}), which gives an AGN incidence of 32\% in this subsample, as opposed to 16\% in the full sample.

\bibliography{biblio}
\bibliographystyle{aasjournal}



\end{document}